\documentclass[11pt]{article}

\usepackage[utf8]{inputenc}
\usepackage{subfigure}
\usepackage{graphicx}
\usepackage{amsmath}
\usepackage{authblk}
\usepackage{fullpage}
\usepackage[sort, round]{natbib}
\usepackage{upgreek}
\usepackage{overpic}
\usepackage[]{multirow}
\usepackage[]{url}
\usepackage{nicefrac}

\begin{document}

\newcommand\Rey{\mbox{\textit{Re}}}  
\newtheorem{lemma}{Lemma}
\newtheorem{corollary}{Corollary}

\title{Constrained Sparse Galerkin Regression}

\author[1]{J.-Ch. Loiseau}
\author[2]{S. L. Brunton}

\affil[1]{Laboratoire DynFluid, Arts et M\'etiers ParisTech, 75013 Paris, France}
\affil[2]{Department of Mechanical Engineering, University of Washington, Seattle, WA 98195, USA}

\date{}

\maketitle

\begin{abstract}

Although major advances have been achieved over the past decades for the reduction and identification of linear systems, deriving nonlinear low-order models still is a challenging task. 
In this work, we develop a new data-driven framework to identify nonlinear reduced-order models of a fluid by combining dimensionality reductions techniques ({\it e.g.} proper orthogonal decomposition) and sparse regression techniques from machine learning. In particular, we extend the sparse identification of nonlinear dynamics (SINDy) algorithm to enforce physical constraints in the regression, namely energy-preserving quadratic nonlinearities. The resulting models, hereafter referred to as \emph{Galerkin regression} models, incorporate many beneficial aspects of Galerkin projection, but without the need for a full-order or high-fidelity solver to project the Navier-Stokes equations. Instead, the most parsimonious nonlinear model is determined that is consistent with observed measurement data and satisfies necessary constraints. Galerkin regression models also readily generalize to include higher-order nonlinear terms that model the effect of truncated modes. The effectiveness of Galerkin regression is demonstrated on two different flow configurations: the two-dimensional flow past a circular cylinder and the shear-driven cavity flow. For both cases, the accuracy of the identified models compare favorably against reduced-order models obtained from a standard Galerkin projection procedure. Present results highlight the importance of cubic nonlinearities in the construction of accurate nonlinear low-dimensional approximations of the flow systems, something which cannot be readily obtained using a standard Galerkin projection of the Navier-Stokes equations. Finally, the entire code base for our constrained sparse Galerkin regression algorithm is freely available online.  

\end{abstract}

\section{Introduction}
\label{sec: introduction}

Fluid flows are characterised by high-dimensional, nonlinear dynamics that give rise to rich structures. Despite this apparent complexity, the dynamics often evolve on a low-dimensional attractor defined by a few dominant coherent structures that contain significant energy or are useful for control \citep{HLBR_turb}. Given this property, one might then aim to derive or identify reduced-order models that reproduce qualitatively and quantitatively the dynamics of the full system. Over the past decades, identifying robust, accurate and efficient reduced-order models has thus become a central challenge in fluid dynamics and closed-loop flow control \citep{fabbiane2014amr, Brunton2015amr, Sipp2016amr, arfm:rowley:2016}.

\bigskip

Many traditional model reduction techniques are analytical. They rely on prior knowledge of the Navier-Stokes equations and the existence of a high-fidelity solver to project onto an orthogonal basis of modes, resulting in a dynamical system in terms of the coefficients of this expansion basis. These modes may come from a classical expansion, such as Fourier modes, or they may be data-driven, as in the proper orthogonal decomposition (POD) \citep{qam:sirovich:1987, berkooz1993proper}. In the latter case, the model-reduction may be considered a hybrid approach, mixing knowledge of the physics with empirical modes obtained from measurement data. Control-theoretic extensions, such as balanced POD (BPOD)~\citep{Willcox2002aiaaj,Rowley2005ijbc}, have also been widely applied for closed-loop flow control~\citep{Ilak2008pof, Bagheri2009jfm, illingworth:2010}. Although such approaches to model reduction have been widely successful for linear systems, as described in the recent review by \cite{arfm:rowley:2016} and references therein, they have been applied with only limited success to obtain low-order approximations of nonlinear systems, mostly on flow oscillators. One can cite for instance the seminal work of \cite{jfm:noack:2003} and \cite{pof:tadmor:2010} wherein the authors have shown that such reduced-order models obtained from a Galerkin projection can reproduce the transients and non-linear dynamics of the von K\`arm\`an vortex shedding past a two-dimensional cylinder provided the projection basis includes a \emph{shift mode} quantifying the distortion between the linearly unstable base flow and marginally stable mean flow. Recently, \cite{jfm:semaan:2016} have extended the reduced-order modeling strategy of~\cite{jfm:noack:2003} to include the effect of control actuation for the flow around a high-lift configuration airfoil.

\bigskip

In contrast, data-driven approaches are becoming increasingly popular and encompass a large variety of different techniques such as the eigensystem realisation algorithm (ERA)~\citep{jgcd:juang:1985}, dynamic mode decomposition (DMD)~\citep{jfm:schmid:2010,Rowley2009jfm,Kutz2016book}, Koopman theory~\citep{Mezic2005nd,Mezic2013arfm} and variants~\citep{Tu2014jcd,Williams2015jnls}, cluster reduced order modeling (CROM)~\citep{Kaiser2014jfm}, and network analysis of fluids~\citep{nair2015network}. Recent advances in machine learning are also greatly expanding the ability to extract governing dynamics purely from data. In particular advanced regression methods from statistics, such as genetic programming  or sparse regression, are driving new algorithms that identify nonlinear dynamics from measurements of complex systems. \cite{Bongard2007pnas} and \cite{Schmidt2009science} introduced nonlinear system identification based on genetic programming, which has been used in numerous practical applications in aerospace engineering, the petroleum industry, and in finance. More recently, \cite{pnas:brunton:2016} have proposed a system identification approach based on sparse regression known as the sparse identification of nonlinear dynamics (SINDy). Following the principle of Ockham's razor, the SINDy algorithm rests on the assumption that there are only a few important terms that govern the dynamics of a given system, so that the equations are sparse in the space of possible functions. Sparse regression is then used to determine the fewest terms in a dynamical system required to accurately represent the data. The resulting models are parsimonious, balancing model complexity with descriptive power while avoiding overfitting.

\bigskip

Most of these regression techniques can be recast into a convex minimisation problem and their solution can be obtained using a number of efficient algorithms available in different libraries such as CVXOPT \citep{CVXOPT}. However, a major drawback of regression-based methods is the possible loss of existing symmetries in the governing equations which may otherwise be included in the physics-based Galerkin projection methods described previously \citep{balajewicz2013low,Carlberg2015siamjsc}. A notable exception is the physics-constrained multi-level quadratic regression used to identify models in climate and turbulence~\citep{Majda2012nonlinearity}. Starting from the original SINDy algorithm \citep{pnas:brunton:2016}, a system identification technique based on sparse regression, we propose in this work a new implementation of the algorithm which allows the user to include physical constraints such as energy-preserving nonlinearities or to enforce symmetries in the identified equations. The resulting algorithm relies on the use of constrained least squares~\citep{book:golub:2012} to incorporate additional constraints in the SINDy algorithm for the sparse identification of the underlying low-dimensional dynamical system. The ability of the present system identification technique, hereafter named {\it sparse Galerkin regression}, is demonstrated on two different flow configurations, namely the emblematic two-dimensional cylinder flow and the shear-driven cavity flow. The manuscript is organised as follows: \textsection \ref{sec: sparse identification}.1 provides the reader with a quick introduction to the original SINDy algorithm, while the new algorithm is presented in \textsection \ref{sec: sparse identification}.2 and illustrated on a toy model in \textsection \ref{sec: sparse identification}.3. The physical constraints used in this work are discussed in \textsection \ref{sec: constraints}, while the two flow configurations considered herein are presented in \textsection \ref{sec: flow configurations}. The different low-dimensional systems identified are compared against standard Galerkin projection in \textsection \ref{sec: discussion}. Finally, \textsection \ref{sec: conclusion} summarises our key findings and provide the reader with possible extensions to this work.

\section{Constrained sparse identification}
\label{sec: sparse identification}

Here we discuss the core mathematical and algorithmic framework used to identify nonlinear reduced-order models from data.  The proposed Galerkin regression method is based on a heavily modified version of the sparse identification of nonlinear dynamics (SINDy) method~\citep{pnas:brunton:2016}. The original SINDy algorithm is introduced in \textsection~\ref{subsec: sindy}, and the new modifications to include physical constraints, such as energy conservation, known eigenvalues, or symmetries, are discussed in \textsection~\ref{subsec: constrained identification}. Implementation details for both algorithms are presented to promote reproducibility; in addition, code is freely available online (\url{https://github.com/loiseaujc/SINDy}). Finally, the inclusion of constraints is demonstrated on the Lorenz system as an illustrative example in \textsection~\ref{subsec: Lorenz system}. Specific constraints that are used to enforce energy conservation are derived later in \textsection~\ref{sec: constraints}.  

	\subsection{Sparse identification of nonlinear dynamics (SINDy)}
	\label{subsec: sindy}
	
	Identifying dynamical systems models from data has been a central challenge in mathematical physics, with a particularly rich history in fluid dynamics. Typically, the form of the dynamical systems model identified is either constrained via prior knowledge of the governing equations, as in Galerkin projection, or a small handful of heuristic models are posited and parameters are optimized to match the data. Simultaneously identifying the structure and parameters of a model from data is considerably more challenging, as there are combinatorially many possible model structures.  
	
	The sparse identification of nonlinear dynamics (SINDy) algorithm~\citep{pnas:brunton:2016} bypasses the intractable brute force search through all possible model structures, leveraging the observation that many dynamical systems
	\begin{equation}
	{\bf \dot{x}} = {\bf f}({\bf x})\label{Eq:Dynamics}
	\end{equation}
 	\noindent have dynamics ${\bf f}$ that are sparse in the space of possible right-hand side functions. It is then possible to solve for the relevant terms that are active in the dynamics using a convex $\ell_1$-regularized regression that penalizes the number of terms in the dynamics and scales well to large problems.  
	
	First, time-series data is collected from Eq.~\eqref{Eq:Dynamics} and formed into a data matrix:
	\begin{equation}
	{\bf X} = \begin{bmatrix} {\bf x}(t_1) & {\bf x}(t_2) & \cdots {\bf x}(t_m)\end{bmatrix}^T
	\end{equation}
	\noindent where $^T$ denotes the matrix transpose.  A similar matrix of derivatives is formed:
	\begin{equation}
	\dot{{\bf X}} = \begin{bmatrix} \dot{{\bf x}}(t_1) & \dot{{\bf x}}(t_2) & \cdots \dot{{\bf x}}(t_m)\end{bmatrix}^T.
	\label{Eq:Xdot}
	\end{equation}
	\noindent In practice, this may be computed directly from the data in ${\bf X}$; for noisy data, the total-variation regularized derivative tends to provide numerically robust derivatives~\citep{Chartrand2011isrnam}.
	
	Based on the data in ${\bf X}$, a library of candidate nonlinear functions $\boldsymbol{\Uptheta}({\bf X})$ is constructed:
	\begin{eqnarray}
	\boldsymbol{\Uptheta}({\bf X}) = \begin{bmatrix} \mathbf{1} & {\bf X} & {\bf X}^2 & \cdots & {\bf X}^d  & \cdots &   \sin({\bf X}) & \cdots  \end{bmatrix}.
	\label{Eq:NLLibrary}
	\end{eqnarray}
	Here, the matrix ${\bf X}^d$ denotes a matrix with column vectors given by all possible time-series of $d$-th degree polynomials in the state ${\bf x}$.  

	The dynamical system in Eq.~\eqref{Eq:Dynamics} may now be represented in terms of the data matrices in Eqs.~\eqref{Eq:Xdot} and~\eqref{Eq:NLLibrary} as
	\begin{eqnarray}
	\dot{{\bf X}} = \boldsymbol{\Uptheta}({\bf X})\boldsymbol{\Xi}.\label{Eq:SINDy1}
	\end{eqnarray}	
	Each column $\boldsymbol{\Xi}_k$ in $\boldsymbol{\Xi}$ is a vector of coefficients determining the active  terms in the $k$-th row equation in Eq.~\eqref{Eq:Dynamics}.  
	A parsimonious model will provide an accurate model fit in Eq.~\eqref{Eq:SINDy1} with as few terms as possible in $\boldsymbol{\Xi}$.  
	Such a model may be identified using a convex $\ell_1$-regularized sparse regression:
	\begin{eqnarray}
	\boldsymbol{\Xi}_k = \text{argmin}_{\boldsymbol{\Xi}_k'}\|\dot{\mathbf{X}}_k - \boldsymbol{\Uptheta}(\mathbf{X})\boldsymbol{\Xi}_k'\|_2+\lambda \|\boldsymbol{\Xi}_k'\|_1.
	\end{eqnarray}
	Here, $\dot{\mathbf{X}}_k$ is the $k$-th column of $\dot{\mathbf{X}}$.  
	Sparse regression, such as the LASSO~\citep{Tibshirani1996lasso} or the sequential thresholded least-squares algorithm used in SINDy, improves the numerical robustness of this identification for noisy overdetermined problems, in contrast to earlier methods~\citep{Wang2011prl} that used compressed sensing~\citep{Donoho2006ieeetit,Candes2006picm}. 
	
	The sparse vectors $\boldsymbol{\Xi}_k$ may be synthesized into a nonlinear dynamical system model: 
\begin{eqnarray}
\dot{x}_k = \boldsymbol{\Uptheta}(\mathbf{x})\boldsymbol{\Xi}_k.
\end{eqnarray}
Note that $x_k$ is the $k$-th element of $\mathbf{x}$ and $\boldsymbol{\Uptheta}(\mathbf{x})$ is a row vector of symbolic functions of $\mathbf{x}$, as opposed to the data matrix $\boldsymbol{\Uptheta}(\mathbf{X})$.

Identifying the most parsimonious nonlinear model by applying sparse regression in the library $\boldsymbol{\Uptheta}$ is a convex procedure.  
The alternative approach, which involves regression onto every possible sparse nonlinear structure, constitutes an intractable brute-force procedure.  
SINDy bypasses this combinatorial search with modern convex optimization and machine learning.  
It is interesting to note that if $\boldsymbol{\Uptheta}({\bf X})$ consists only of linear terms, and if we remove the sparsity promoting term by setting $\lambda=0$, then this algorithm reduces to the dynamic mode decomposition~~\cite{jfm:schmid:2010,Rowley2009jfm,Kutz2016book}.  
		
	Recent extension to SINDy enable the identification of nonlinear differential equations with rational function nonlinearities by reformulating the problem as an implicit differential equation and solving for the active terms by finding the sparsest vector in the null space of an augmented library containing functions of the state and derivative terms~\citep{Mangan2016ieee}.  
	SINDy has also been generalized to identify partial differential equations from data~\citep{Rudy2016arxiv}, and has been extended to include inputs and control~\citep{Brunton2016nolcos}. 
	
	\subsection{Constrained sparse identification}
	\label{subsec: constrained identification}
	
	It has been shown in \textsection \ref{subsec: sindy} that, within the SINDy framework, the identification problem can be cast as a convex optimisation problem where the sparsity of the solution ${\boldsymbol{\Xi}}$ can be promoted using an $l_1$ regularized regression. Alternatively, sparsity can also be promoted by using the sequential thresholded least-squares algorithm as in~\cite{pnas:brunton:2016}. In this case, the convex minimisation problem can be re-written as
	
	\begin{equation}
	    \begin{aligned}
    	    & \min_{\boldsymbol{\Xi}} \Vert {\boldsymbol{\Uptheta}}({\bf X}) {\boldsymbol{\Xi}} - \dot{\bf X} \Vert_2^2 \\
    	    & \text{subject to } {\bf C}{\boldsymbol \upxi} = {\bf d}
    	\end{aligned}
	    \label{eq: sequential thresholded least squares}
	\end{equation}
	
	\noindent where ${\boldsymbol \upxi}=\boldsymbol{\Xi}(:)$ is the vectorized form of the sparse matrix of coefficients, and where ${\bf C}{\boldsymbol \upxi} = {\bf d}$ are linear equality constraints, which can be used to enforce that some entries of ${\boldsymbol \upxi}$ are equal to zero. The minimisation problem is then solved iteratively. After an initial least-squares regression, the thresholding is performed as follows: if $\lvert \upxi_i \rvert$ is smaller than $\lambda$ (the sparsity knob) times the mean of the absolute value of the non-zero entries of $ {\boldsymbol \upxi} $, then an additional row is added to the constraint matrix ${\bf C}$ to enforce $\upxi_i=0$. Two or three iterations of this small variation of the sequential thresholded least-squares algorithm are usually sufficient to ensure convergence of the constrained minimization procedure. 
	The sparsity parameter $\lambda$ should be chosen to promote parsimonious models that strike a balance between accuracy and complexity to avoid overfitting the data.  More details on this choice are presented in Appendix~\ref{AppendixB}.
	
	\bigskip
	
	From a practical point of view, each iteration of \eqref{eq: sequential thresholded least squares} can be recast as an unconstrained problem by using an augmented functional formulation where the constraints are imposed by means of Lagrange multipliers. The resulting unconstrained minimisation problem then reads
	
	\begin{equation}
	    \min_{{\boldsymbol \upxi}, {\bf z}} \Vert {\boldsymbol{\Uptheta}}({\bf X}) {\boldsymbol{\Xi}} - \dot{\bf X} \Vert_2^2 + {\bf z}^T({\bf C} {\boldsymbol \upxi} - {\bf d}).
	\end{equation}
	
	\noindent Given the choice of our augmented functional, it can easily be shown that the optimal solution ${\boldsymbol \upxi}$ that satisfies the constraints is also solution to the following Karush-Kuhn-Tucker (KKT) equations
	
	\begin{equation}
	    \begin{bmatrix}
	    2 \boldsymbol{\hat\Uptheta}({\bf X})^T \boldsymbol{\hat\Uptheta}({\bf X}) & {\bf C}^T \\
	    {\bf C} & 0
	    \end{bmatrix} \begin{bmatrix}
	                    {\boldsymbol \upxi} \\
	                    {\bf z}
	                  \end{bmatrix} = \begin{bmatrix}
	                                    2\boldsymbol{\hat\Uptheta}({\bf X})^T \dot{\bf X}(:) \\
	                                    {\bf d}
	                                  \end{bmatrix},
	   \label{eq: KKT equations}
	\end{equation}
	
	\noindent where $\boldsymbol{\hat\Uptheta}({\bf X})$ is a diagonal matrix consisting of $n$ copies of $\boldsymbol{\Uptheta}({\bf X})$, ${\bf X}(:)$ is the vectorized form of ${\bf X}$ (same as the vectorization of $\boldsymbol{\Xi}$ into ${\boldsymbol \upxi}=\boldsymbol{\Xi}(:)$), and $n$ is the dimension of ${\bf x}$. This matrix equation for constrained least-squares is the counterpart to the ordinary least-squares normal equations. It has a unique solution if ${\bf C}$ has full row-rank and $\begin{bmatrix} \boldsymbol{\hat\Uptheta}({\bf X}) & {\bf C} \end{bmatrix}^T$ has full column-rank.

	\medskip    
    
    Interestingly, the linear equality constraints ${\bf C} {\boldsymbol \upxi} = {\bf d}$ do not have to be used for the sole purpose of sparsity promotion. Indeed, these can also be used to enforce additional user-provided constraints such as an {\it a priori} known value of a given entry $\upxi_i$ or to impose some linear relationship between the entries of ${\boldsymbol \upxi}$ to mimic a given physical process, see \textsection \ref{subsec: Lorenz system} for a simple illustration. Specific constraints required to conserve energy in a fluid are derived later in \textsection~\ref{sec: constraints}.  	
		
	\subsection{Illustration of constrained sparse identification on the Lorenz system}
	\label{subsec: Lorenz system}
	
	Following \cite{pnas:brunton:2016}, let us first illustrate how to formulate user-provided constraints using the Lorenz system \citep{Lorenz1963jas}. This dynamical system, derived by Edward Lorenz in 1963, is notable for having chaotic solutions for certain parameter values and initial conditions. It reads
	
	\begin{equation}
	    \begin{aligned}
	        & \dot{x} = \sigma ( y - x ) \\
	        & \dot{y} = x ( \rho - z ) - y \\
	        & \dot{z} = xy - \beta z.
        \end{aligned}
        \label{eq: Lorenz system}
	\end{equation}
	
\begin{figure}
	    \includegraphics[width=\textwidth]{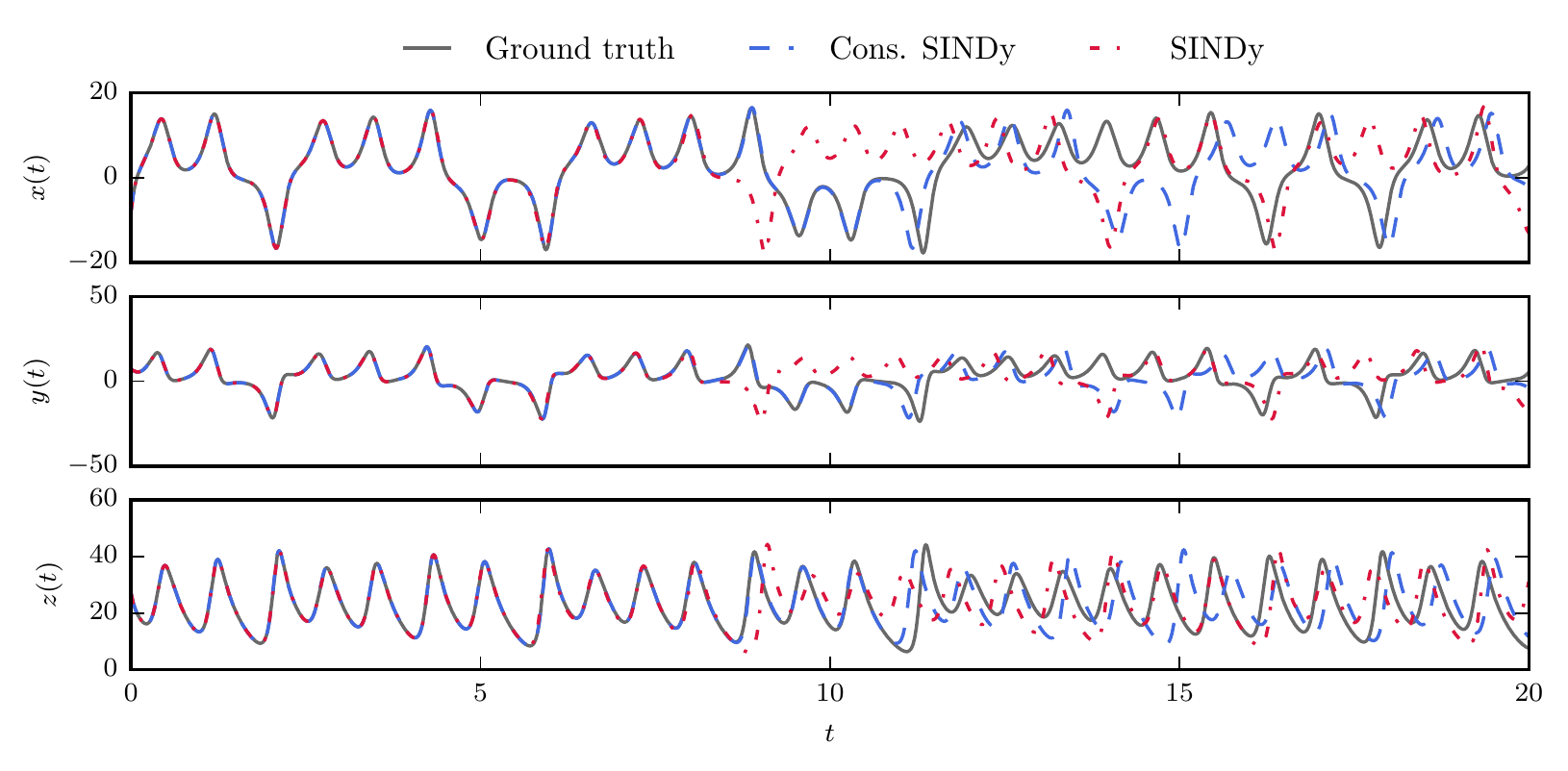}
	    \caption{Dataset (grey) used for the constrained sparse identification of the Lorenz system and prediction obtained using the model identified by the constrained SINDy algorithm (blue dashed) and the original SINDy algorithm (red dashed).}
	    \label{fig: Lorenz system}
	\end{figure}

	\noindent Figure \ref{fig: Lorenz system} depicts the evolution in time of ${\bf x}(t) = [ x(t), \ y(t), \ z(t) ]^T$ for a given set of parameters $\sigma$, $\rho$ and $\beta$. These signals, as well as their derivatives (not shown), will serve as the input data for the constrained system identification. For that purpose, the library $\boldsymbol{\Uptheta} ({\bf x})$ used in the identification process is defined as $P_2({\bf x})$, {\it i.e.} all the polynomials of degree 2 or less in the entries of ${\bf x}$ such that
	
	\begin{equation}
	    \boldsymbol{\Uptheta}({\bf x}) = \begin{bmatrix}
	                        1 & x & y & z & x^2 & xy & xz & y^2 & yz & z^2
	                        \end{bmatrix}.
	\end{equation}
	
	\noindent Up to 30 different coefficients thus need to be identified, 10 per equation. Let us assume furthermore that, in the $x$-equation, we know beforehand that $\sigma=10$. The constrained optimisation problem on which SINDY relies then reads
	
	\begin{equation}
        \begin{aligned}
            & \min_{{\boldsymbol{\Xi}}} \| \dot{{\bf X}} - \boldsymbol{\Uptheta}({\bf X}) {\boldsymbol \upxi} \|^2_2 \\
            \text{subject to } & \xi_3 = 10 \\
                              & \xi_2 + \xi_3 = 0.
        \end{aligned}
        \label{eq: constrained regression problem}
    \end{equation}
    
    \noindent From a practical point of view, these equality constraints are passed to CVXOPT as ${\bf C} {\boldsymbol \upxi} = {\bf d}$, where ${\bf C}$ is a $2 \times 30$ matrix and ${\bf d}$ a vector given by
    
    \begin{equation}
        {\bf C} = \begin{bmatrix}
                  0 & 0 & 1 & 0 & \cdots & 0 \\
                  0 & 1 & 1 & 0 & \cdots & 0 
                  \end{bmatrix} \text{ and } {\bf d} = [10 \ 0]^T.
    \end{equation}
    
    \noindent Using a suitable sparsity knob, the system identified by the constrained SINDy algorithm finally reads
    
	\begin{equation}
	    \begin{aligned}
	        & \dot{x} = 10 ( y - x ) \\
	        & \dot{y} = x (27.99 - 0.999z ) - 0.998y \\
	        & \dot{z} = 0.999xy - 2.666 z.
        \end{aligned}
        \label{eq: Identified Lorenz system}
	\end{equation}
	
	\noindent The coefficients of the identified system are close to the original ones, which were set to $\sigma = 10$, $\rho = 28$ and $\beta = \nicefrac{8}{3}$. The time-evolution given by this identified system is depicted in figure \ref{fig: Lorenz system} along with the original signals and those given by a system identified using the original (unconstrained) SINDy algorithm. It can be seen that the trajectory of the system identified using constrained SINDy remains closer to that of the original system compared to the trajectory predicted by the system identified using the original SINDy. The effects of adding constraints are even more pronounced in fluid systems where energy conservation may be enforced if certain constraints on the quadratic nonlinearities are satisfied, as discussed in \textsection~\ref{sec: constraints} and \textsection~\ref{sec: discussion}.  
	
\section{Deriving the constraints}
\label{sec: constraints}	

	The Navier-Stokes equations governing the dynamics of the perturbation ${\bf u}$ evolving on top of the base flow ${\bf U}_b$ are given by	
	
	\begin{eqnarray}
		& \displaystyle \frac{\partial {\bf u}}{\partial t} = -({\bf U}_b \cdot \nabla ) {\bf u} - ({\bf u} \cdot \nabla ) {\bf U}_b - \nabla p + \frac{1}{\Rey} \nabla^2 {\bf u}   - ({\bf u} \cdot \nabla) {\bf u}\\
		& \nabla \cdot {\bf u} = 0
		\label{eq: perturbative Navier-Stokes equations}
	\end{eqnarray}
	
	\noindent where ${\bf U}_b$ is the base flow velocity field, ${\bf u}$ is the perturbation velocity field and $p$ the corresponding pressure. The aim of reduced-order modeling is to construct/derive/identify a low-dimensional system of the form
	
	\begin{equation}
		\frac{\mathrm{d} {\bf a}}{\mathrm{d}t} = \tilde{\mathcal{L}}{\bf a} + \tilde{\mathcal{N}}({\bf a}){\bf a}
		\label{eq: reduced-order model}
	\end{equation}
	
	\noindent where $\tilde{\mathcal{L}}$ and $\tilde{\mathcal{N}}({\bf a}){\bf a}$ are low-dimensional approximation of the linearised Navier-Stokes operator and of the quadratic nonlinear term, respectively, and where the entries of ${\bf a}$ are the degrees of freedom of the reduced-order model. For the reduced-order model \eqref{eq: reduced-order model} to be a good approximation of its high-dimensional counterpart, the former needs to have the same physical properties as the latter. While this is expected to be true when the reduced-order model is derived based on a Galerkin projection, these properties need to be actively enforced when a system identification approach as SINDy is used.
	
	\subsection{Constraining the quadratic nonlinear term}
	\label{subsec: quadratic constraints}
	
	The nonlinear Navier-Stokes equations \eqref{eq: perturbative Navier-Stokes equations} are partial differential equations characterised by the quadratic nonlinear term $-({\bf u} \cdot \nabla){\bf u}$. It can be shown that
	
	\begin{equation}
		\int_{\Omega} {\bf u} \cdot ({\bf u} \cdot \nabla){\bf u} \, \mathrm{d}\Omega = 0
		\label{eq: energy-preserving nonlinearity}
	\end{equation}
	
	\noindent where the boundary terms resulting from the integration by parts are assumed to be small enough and can thus be neglected for the sake of simplicity. The contribution of the quadratic nonlinear term to the total energy of the perturbation is zero: it is an energy-preserving nonlinearity, its role being only to redistribute the perturbation's energy along the different lengthscales of the problem.
	
	
	 Given that our projection basis contains the POD modes, their amplitudes $a_i(t)$ are directly related to the kinetic energy of the perturbation. The constraint required in our system identification for the low-dimensional quadratic nonlinear term to be energy-preserving is thus	
	\begin{equation}
		{\bf a} \cdot \tilde{\mathcal{N}}({\bf a}){\bf a} = 0.
		\label{eq: quadratic constraint}
	\end{equation}

    \noindent Expanding \eqref{eq: quadratic constraint} in terms of the regression coefficients ${\boldsymbol \upxi}$ yields    
    \begin{equation}
        0 = {\bf a}^T \begin{bmatrix}
        						\upxi_4^{(a_1)} a_1^2 & \upxi_5^{(a_1)} a_1 + \upxi_7^{(a_1)}a_2 & \upxi_6^{(a_1)} a_1 + \upxi_9^{(a_1)} a_{\Delta} \\
        						\upxi_4^{(a_2)} a_1 + \upxi_5^{(a_2)} a_2 & \upxi_7^{(a_2)} a_2 & \upxi_8^{(a_2)} a_2 + \upxi_9^{(a_2)} a_{\Delta} \\
        						\upxi_4^{(a_{\Delta})} a_1 + \upxi_6^{(a_{\Delta})} a_{\Delta} & \upxi_7^{(a_{\Delta})} a_y + \upxi_8^{(a_{\Delta})} a_{\Delta} & \upxi_9^{(a_{\Delta})} a_{\Delta}
        						\end{bmatrix} \begin{bmatrix}
        												a_1 \\
        												a_2 \\
        												a_{\Delta}
        												\end{bmatrix} + {\bf a}^T \begin{bmatrix}
        																							\upxi_8^{(a_1)} a_2 a_{\Delta} \\
        																							\upxi_6^{(a_2)} a_1 a_{\Delta} \\
        																							\upxi_5^{(a_{\Delta})} a_1 a_2
        																							\end{bmatrix}
        \label{eq: quadratic constraint bis}
    \end{equation}
    
   \noindent For  \eqref{eq: quadratic constraint bis} to hold, the matrix involved in the first term is required to be skew-symmetric, while the second term implies $\upxi_8^{(a_1)} + \upxi_6^{(a_2)} + \upxi_5^{(a_{\Delta})} = 0$. Overall, this gives rise to ten different linear equality constraints which induce a coupling of the different ordinary differential equations governing the evolution of $a_1$, $a_2$ and $a_{\Delta}$.
    
	\subsection{What about higher order nonlinearities?}
	\label{subsec: higher-order constraints}
	
	Reduced-order modelling based on Galerkin projection usually requires a relatively large projection basis. Despite the very low effective dimensionality of the cylinder flow at $\Rey=100$, \cite{jfm:noack:2003} demonstrated the need to include the first eight POD modes along with the shift mode for the reduced-order model to provide a relatively faithful approximation of the original high-dimensional dynamics. Including the higher harmonic POD modes was deemed necessary in order to limit the energy overshoot otherwise observed during the nonlinear saturation process. Even though they might be required to prevent a non-physical behaviour of the reduced-order model, these higher harmonic modes have very low energy and limited dynamics of their own: they are essentially {\it enslaved} to the dominant POD modes. Using {\it adiabatic elimination}~\citep{springer:haken:1983} or {\it center manifold reduction}~\citep{book:wiggins:2003, jfm:carini:2015}, it is well known that these slaved modes can be reduced out of the problem, while their influence onto the driving modes can be accounted for by appropriately modifying the nonlinear terms, generally introducing higher-order nonlinearities. Such an approach to reduced-order modelling, which can be summarised as {\it derive-then-reduce}, can be used to reduce the eight-dimensional system derived by \cite{jfm:noack:2003} for the two-dimensional cylinder flow into one having only three degrees of freedom, {\it i.e.} the amplitude of the shift mode and that of the first two POD modes. 
	
	This derive-then-reduce approach is generally quite involved, requiring cumbersome calculations, particularly if the original Galerkin projection model has more than just a few degrees of freedom. However, in the present work, high-order nonlinearities modelling the influence of the truncated modes can be automatically incorporated in the identification process, with no additional post-analysis. For that purpose, the library ${\boldsymbol{\Uptheta}}({\bf a})$ of admissible functions needs to be extended in order to include higher-order polynomials. Note, however, that it is unclear at the present time how to constrain these high-order nonlinearities to ensure that the identified model is physical, although the method is effective in practice without constraining the higher-order terms.  
	
\section{Flow configurations}
\label{sec: flow configurations}
	To demonstrate the Galerkin regression framework, we consider two prototypical flow configurations, the incompressible flow past a circular cylinder and the shear-driven cavity flow.  
	These flows have been selected because they are standard benchmark problems for modal analysis, model reduction, and control in the literature, and because they provide a balance between complexity and interpretability.  
	
	\subsection{Cylinder flow}
	
	The first flow configuration considered in the present work is the two-dimensional incompressible viscous flow past a circular cylinder at ${\Rey = 100}$. This Reynolds number, based on the free-stream velocity $U_{\infty}$, the cylinder diameter $D$ and the kinematic viscosity $\nu$, is well above the onset of vortex shedding~\citep{jem:zebib:1987, jfm:schumm:1994} and below the onset of three-dimensional instabilities~\citep{pof:zhang:1995, jfm:barkley:1996}. In the fluid dynamics community, a large body of literature exists in which this particular setup has been chosen to illustrate modal decomposition~\citep{jfm:bagheri:2013} and model identification techniques~\citep{jfm:noack:2003, pre:sengupta:2015, pnas:brunton:2016, arfm:rowley:2016}. This setup is thus a particularly compelling test case to illustrate our model identification strategy, as well as to draw connections and quantify its performance against other well-established techniques, namely Galerkin projection.
	
	\bigskip
	
\begin{figure}
		\centering
		\subfigure[]{\hspace{-1.1cm} \includegraphics[width=.9\textwidth]{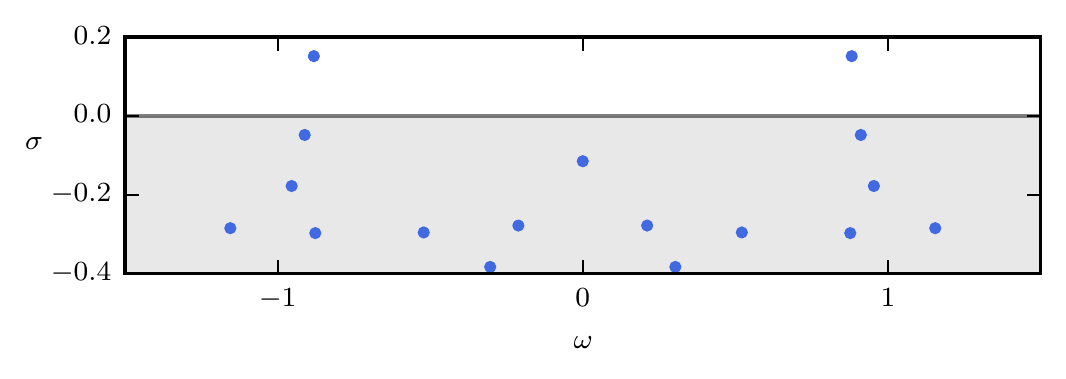}}
		\subfigure[]{\includegraphics[width=.45\textwidth]{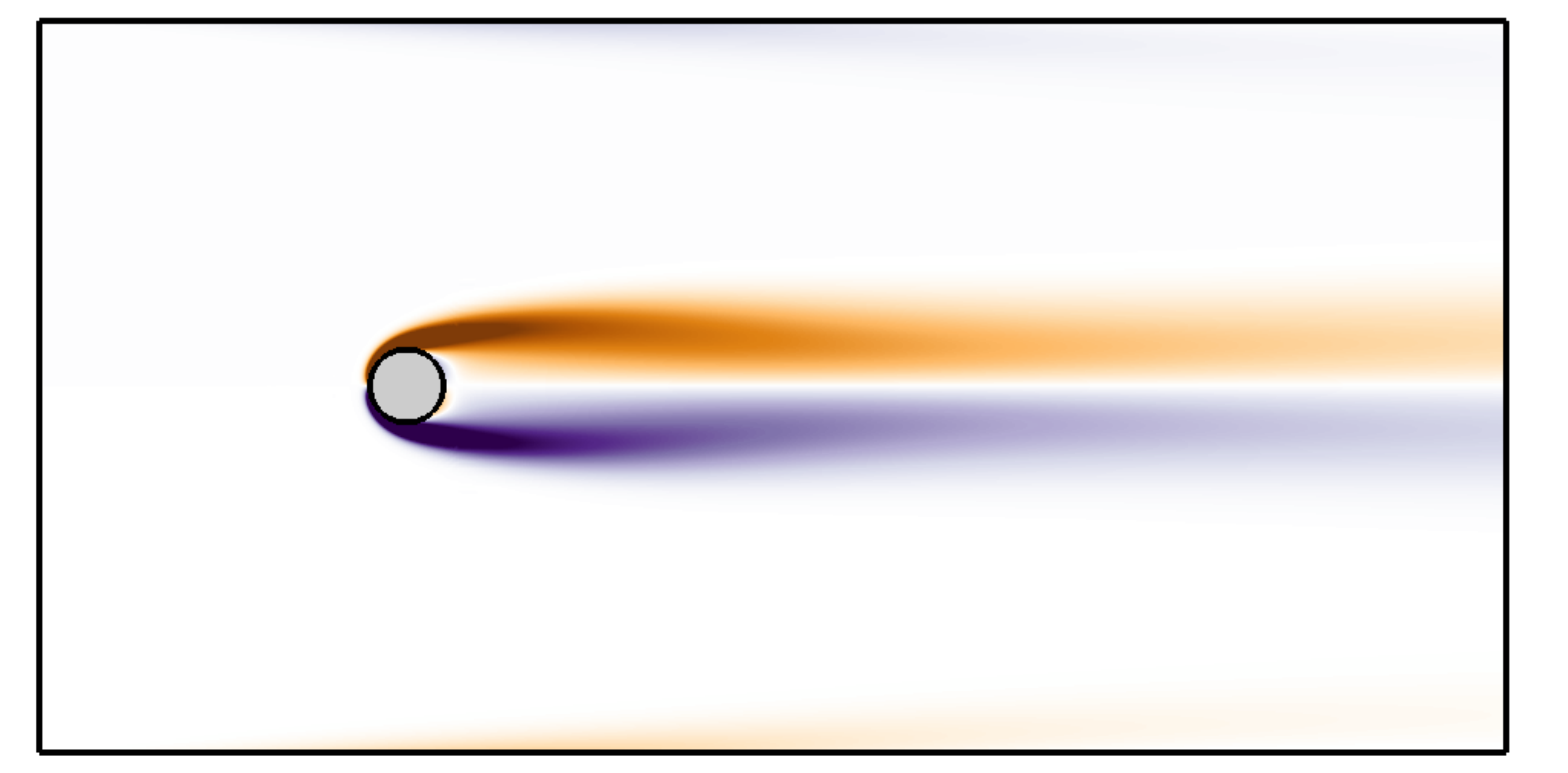}}
		\subfigure[]{\includegraphics[width=.45\textwidth]{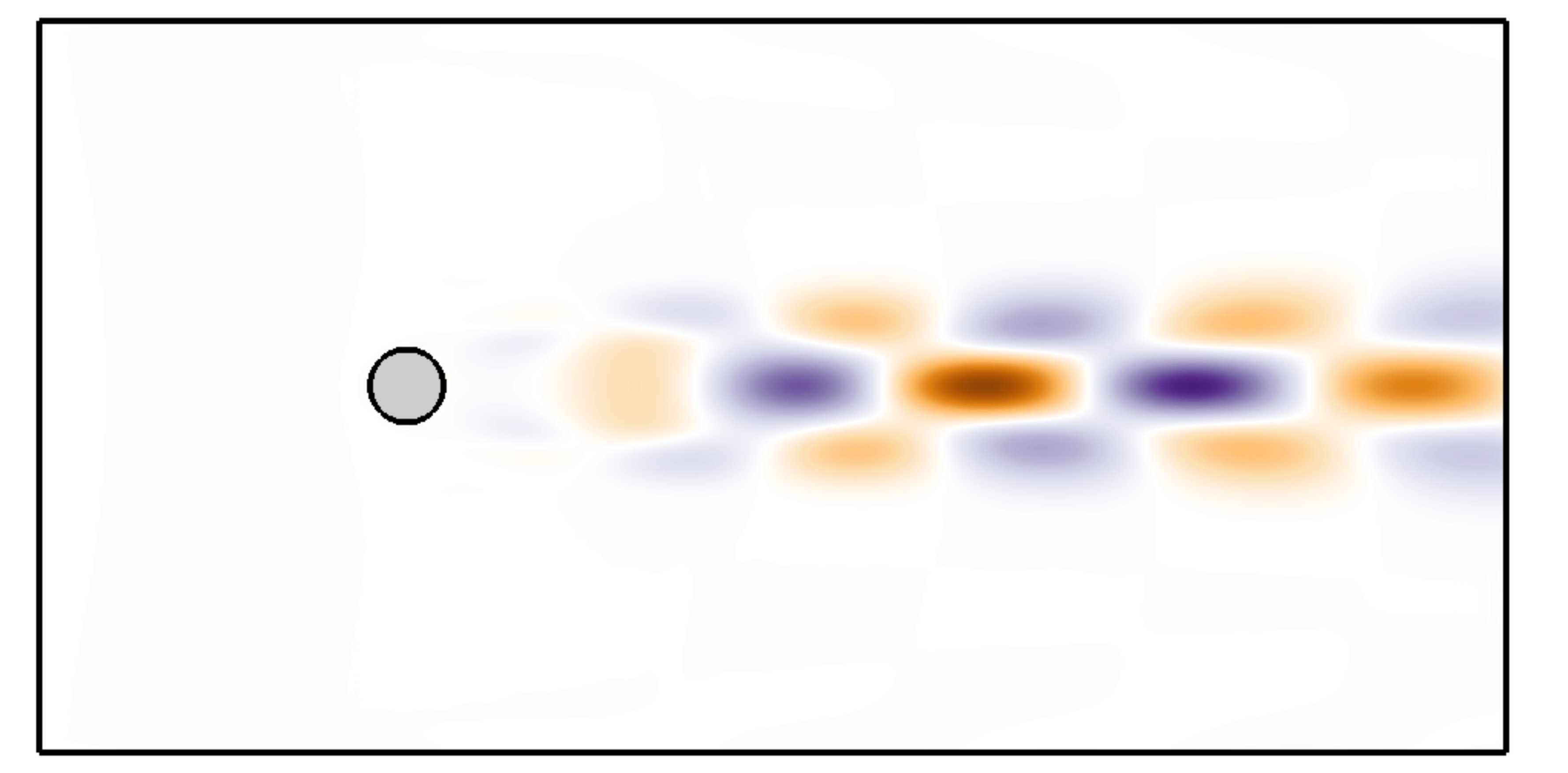}}
		\caption{(a) Eigenspectrum of the linearised Navier-Stokes operator for the two-dimensional cylinder flow at $\Rey=100$. Vorticity fields of (b) the base flow and (c) the leading linearly unstable eigenmode.}
		\label{fig: cylinder stability}
	\end{figure}

	The dynamics of the flow are governed by the incompressible Navier-Stokes equations. These are solved numerically using the Nek 5000 spectral element solver~\citep{nek5000_site}. The same computational domain as in \cite{jfm:noack:2003} has been considered. It extends from $x_1=-5$ up to $x_1=15$ in the streamwise direction, and from $x_2=-5$ up to $x_2=5$ in the spanwise direction. It is discretised using 1832 seventh-order spectral elements. The vorticity field of the linearly unstable fixed point ${\bf U}_b$, computed using the selective frequency damping approach \citep{pof:akervik:2006}, is shown in figure \ref{fig: cylinder stability}(b). Figure \ref{fig: cylinder stability}(a) and (c) also provide the eigenspectrum of the linearised Navier-Stokes operator and the vorticity field associated to the leading unstable eigenmode for the sake of completeness. Though this eigenmode is clearly related to vortex shedding, it is well known that both its spatial distribution and the frequency of the associated eigenvalue differ quite significantly from that of the non-linearly saturated von K\`arm\`an vortex street \citep{epl:barkley:2006}.
	
	\bigskip
	
\begin{figure}
        \centering
        \subfigure[]{\includegraphics[width=.325\columnwidth]{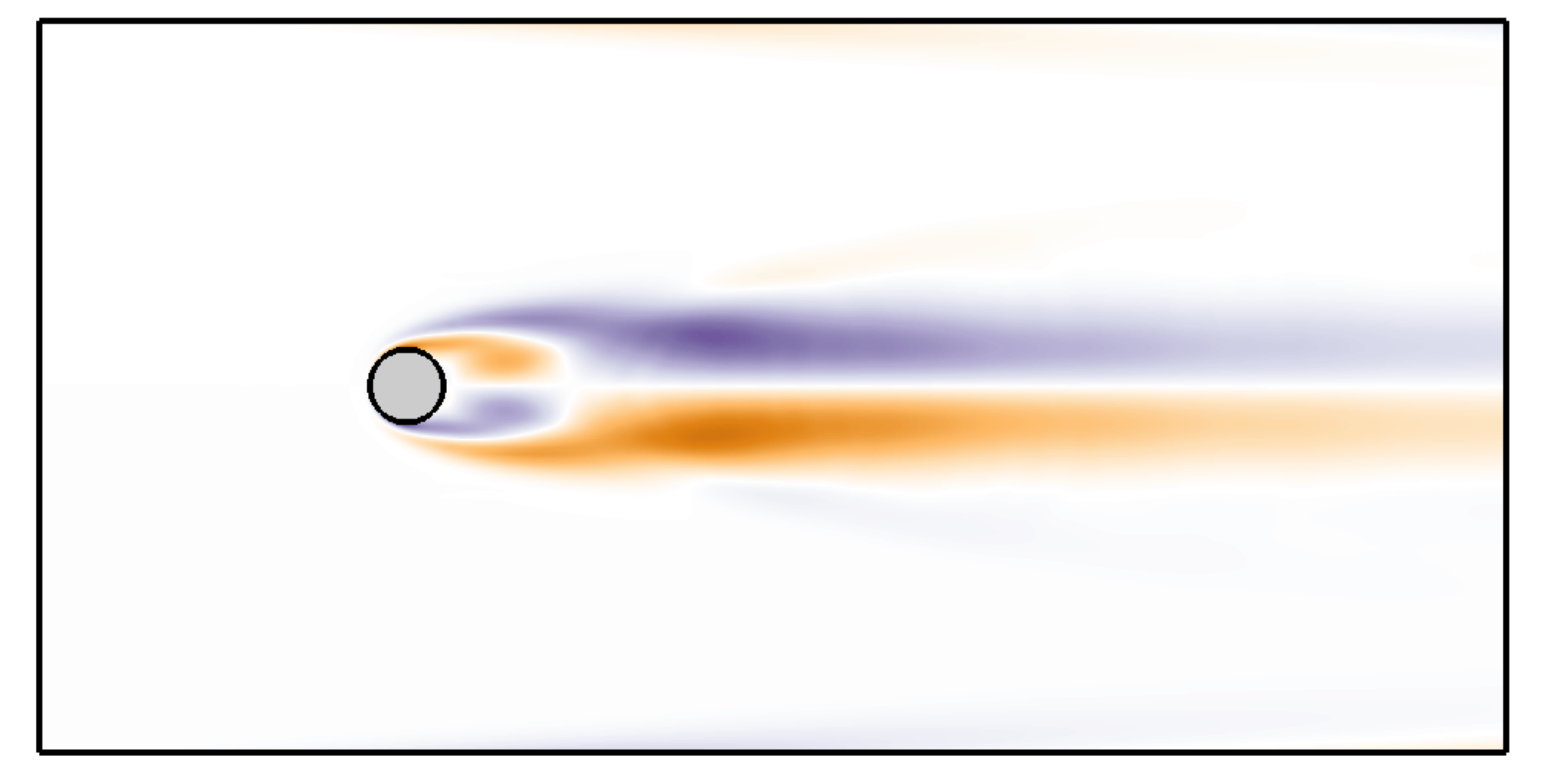}}
        \subfigure[]{\includegraphics[width=.325\columnwidth]{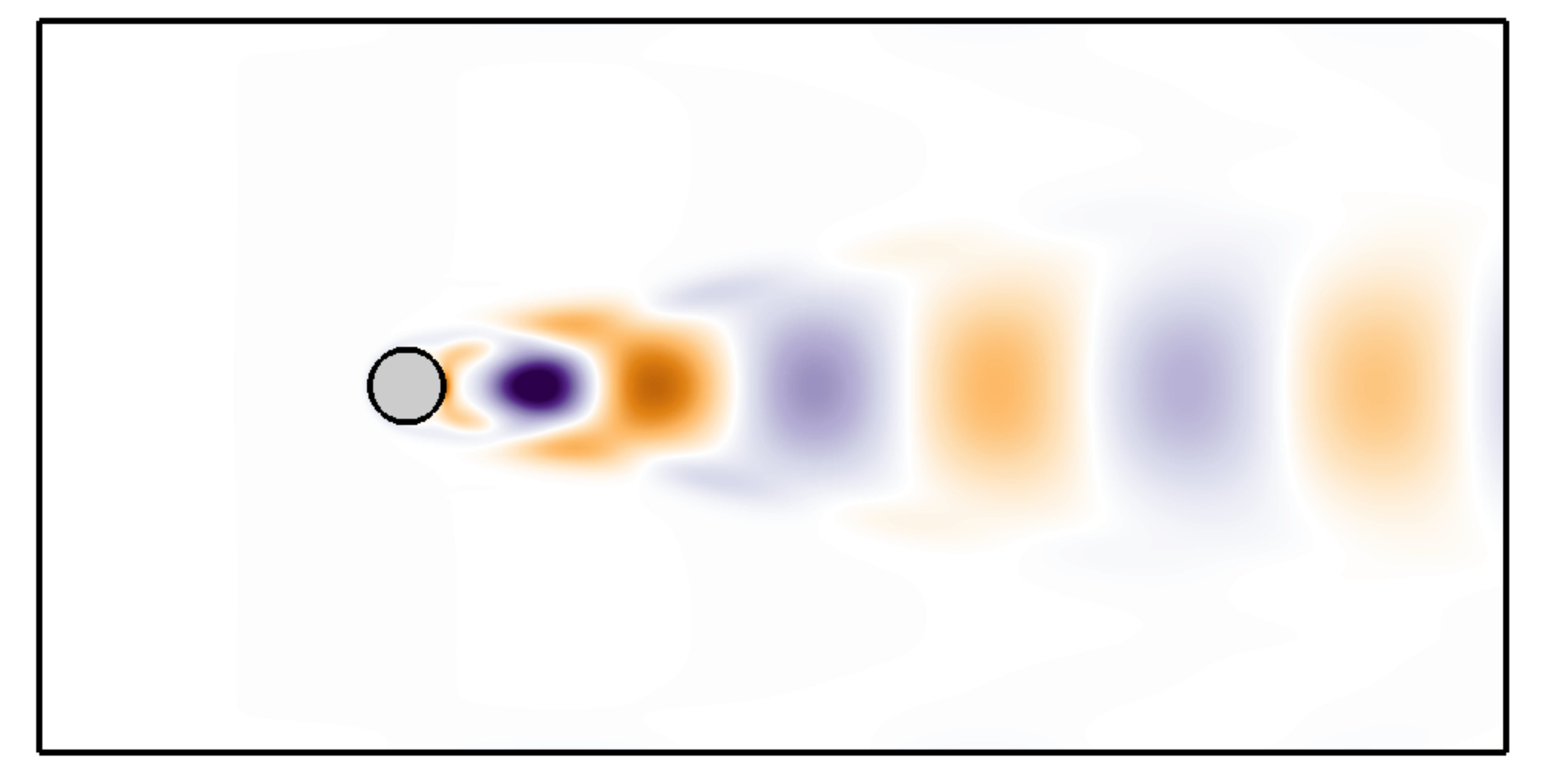}}
        \subfigure[]{\includegraphics[width=.325\columnwidth]{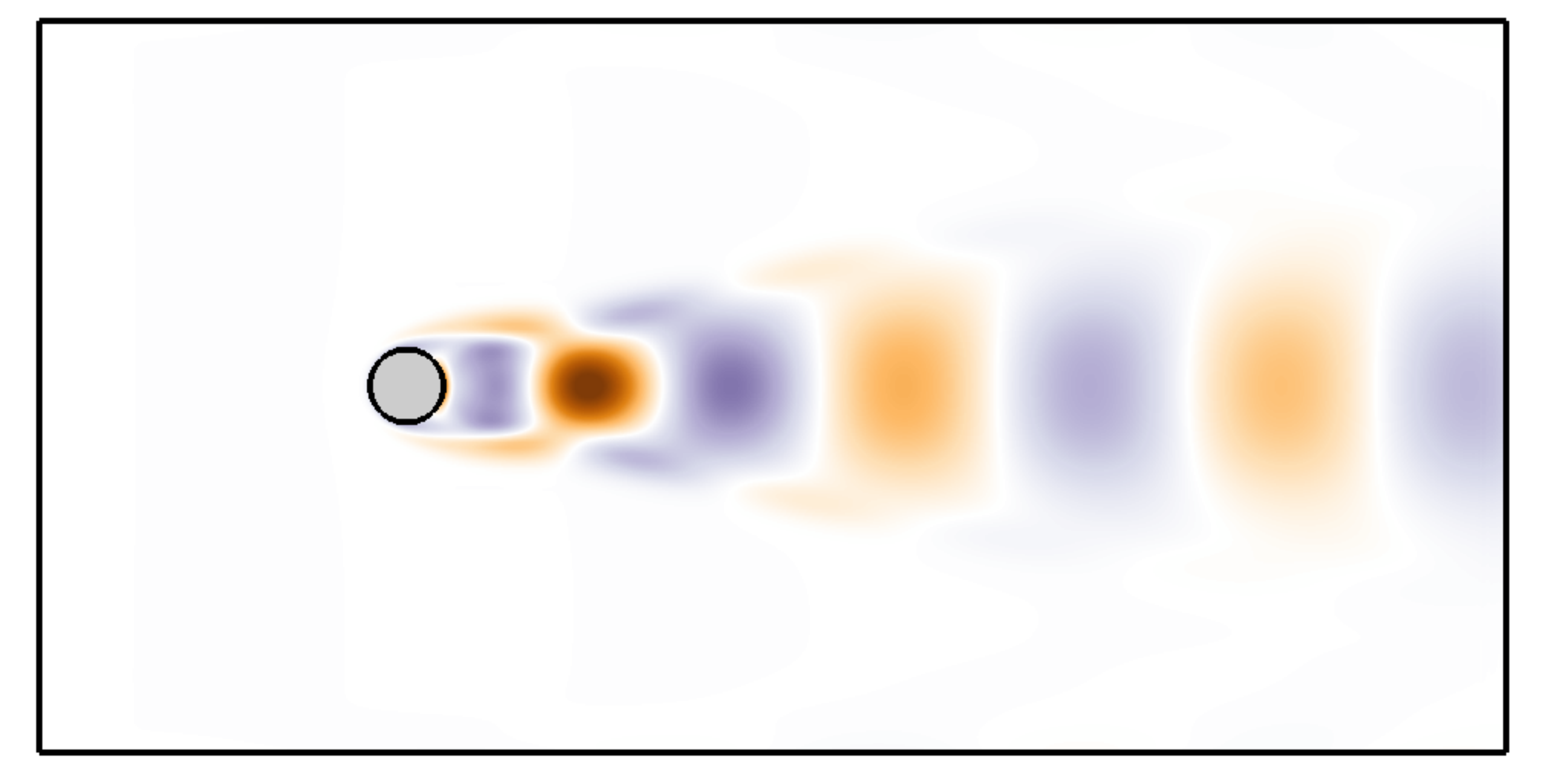}}
        \caption{Vorticity fields of (a) the shift mode, (b) the first and (c) second POD modes of the cylinder flow at $Re=100$.}
        \label{fig: cylinder POD modes}
    \end{figure}

\begin{figure}
        \centering
        \includegraphics[width=\textwidth]{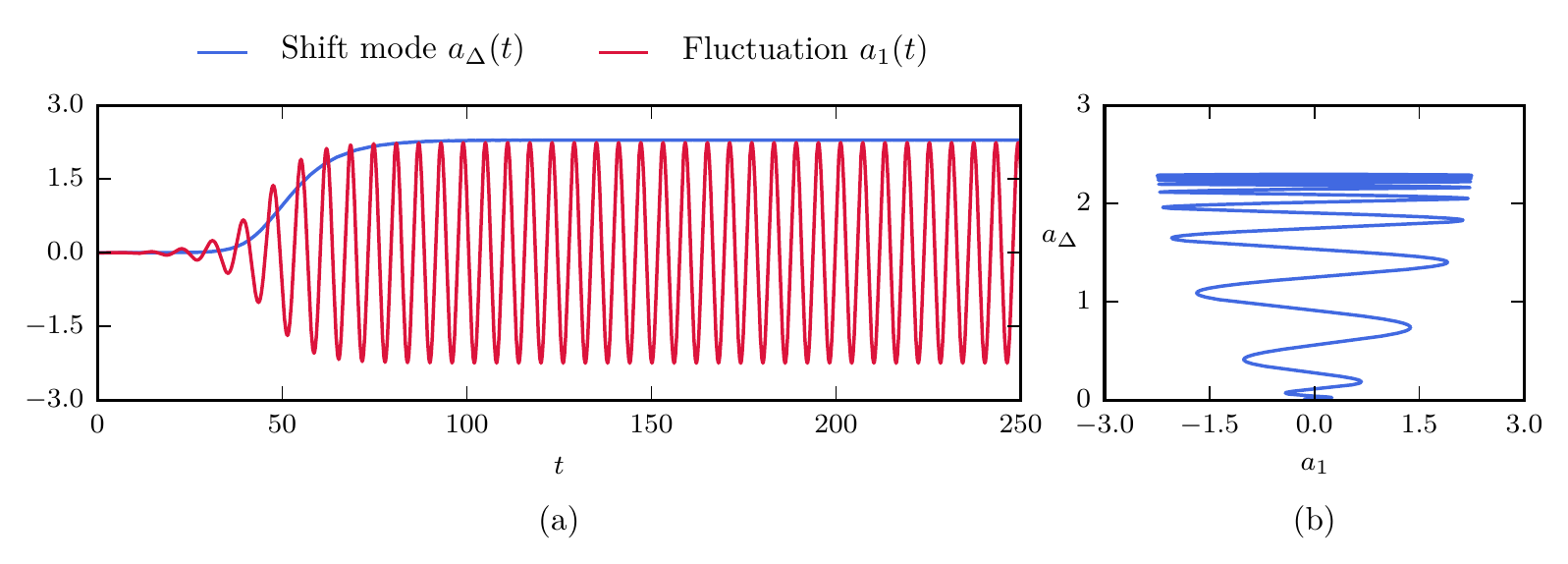}
        \vspace{-.25in}
        \caption{(a) Time evolution of the POD coefficients for the cylinder flow at $\Rey=100$. The time evolution of $a_2(t)$, not shown, is very similar to that of $a_1(t)$. (b) Trajectory in the phase space projected onto the $a_1-a_{\Delta}$ plane.}
        \label{fig: cylinder dataset}
    \end{figure}

	Given this linearly unstable base flow as initial condition, a direct numerical simulation has been run until a statistically steady-state has been achieved. The dynamics of the system on the final attractor are then equidistantly sampled using $M=1000$ velocity field snapshots with a sampling frequency about 30 times larger than the vortex shedding frequency \citep{arxiv:noack:2015}. The shift mode, denoted ${\bf u}_{\Delta}$ and depicted in figure \ref{fig: cylinder POD modes}(a), quantifies the distortion between the unstable base flow equilibrium and the mean flow. It has been shown to be crucially important for POD-based reduced-order modeling~\citep{jfm:noack:2003, pof:tadmor:2010}. The snapshot POD method of \cite{qam:sirovich:1987} has then been used to extract the two most energetic modes ${\bf u}_1$ and ${\bf u}_2$, depicted in figures \ref{fig: cylinder POD modes}(b) and \ref{fig: cylinder POD modes}(c), respectively. The evolution in time of the POD coefficients is shown in figure \ref{fig: cylinder dataset}(a), while a projection of the system's trajectory onto the $a_1-a_{\Delta}$ plane is depicted in figure \ref{fig: cylinder dataset}(b), where $a_1(t)$ is the amplitude of the POD mode ${\bf u}_1$ and $a_{\Delta}(t)$ the amplitude of the shift mode ${\bf u}_{\Delta}$. These signals and their time derivatives (not shown) form the training dataset used to identify the models in \textsection \ref{Results:Cylinder}.
    
	\subsection{Shear-driven cavity flow}
	
    The second flow configuration investigated is the incompressible shear-driven cavity flow. It is a geometrically-induced separated boundary layer flow having a number of applications in aeronautics. The leading two-dimensional instability of the flow is mostly localised along the shear layer developing at the interface between the outer boundary layer flow and the inner cavity flow \citep{amr:sipp:2010}. This oscillatory global instability of the external shear layer relies on two essential mechanisms. On the one hand, the convectively unstable nature of the shear layer causes perturbations to grow as they are convected downstream, while on the other hand, the feedback mechanism provided by the inner-cavity recirculating flow allows these same perturbations to eventually re-excite the upstream shear layer. The coupling between these two mechanisms gives rise to a linearly unstable feedback loop at sufficiently high Reynolds numbers. Note that for compressible shear-driven cavity flows, a similar unstable feedback loop exists wherein the feedback mechanism is provided by upstream-propagating acoustic waves \citep{techreport:rossiter:1964, jfm:rowley:2002, jfm:yamouni:2013}. This strictly two-dimensional linearly unstable flow configuration has served multiple purposes over the past decade: illustration of optimal control and reduced-order modelling \citep{jfm:barbagello:2009}, investigation of the nonlinear saturation process of globally unstable flows \citep{jfm:sipp:2007}, or as an introduction to dynamic mode decomposition \citep{jfm:schmid:2010}, to name just a few.

    \bigskip

\begin{figure}
	\centering
	\subfigure[]{\hspace{-1.1cm} \includegraphics[width=.9\textwidth]{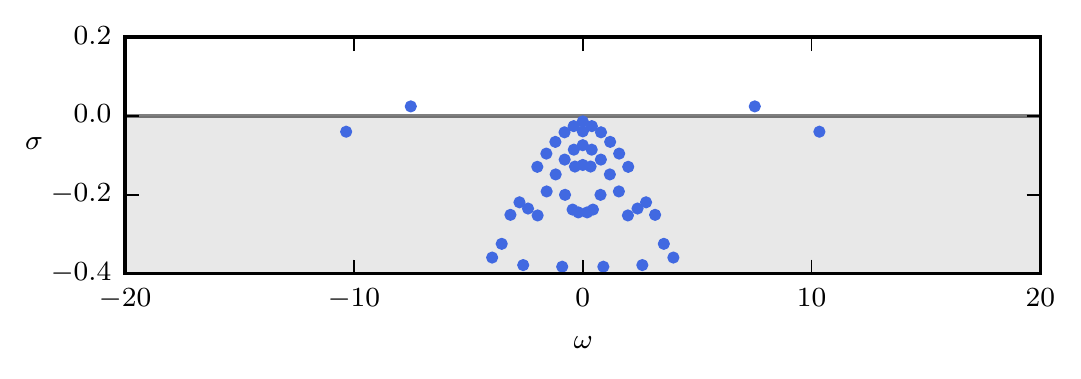}}
	\subfigure[]{\includegraphics[width=.45\textwidth]{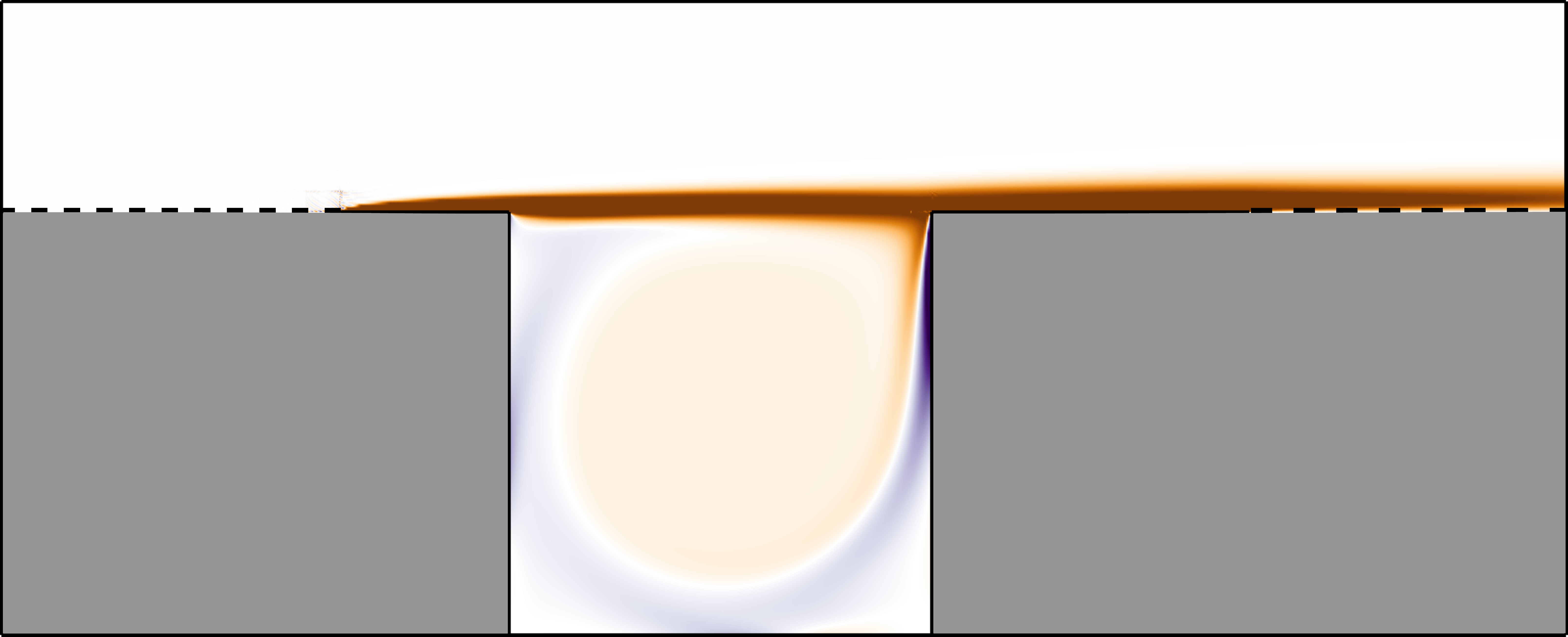}}
	\subfigure[]{\includegraphics[width=.45\textwidth]{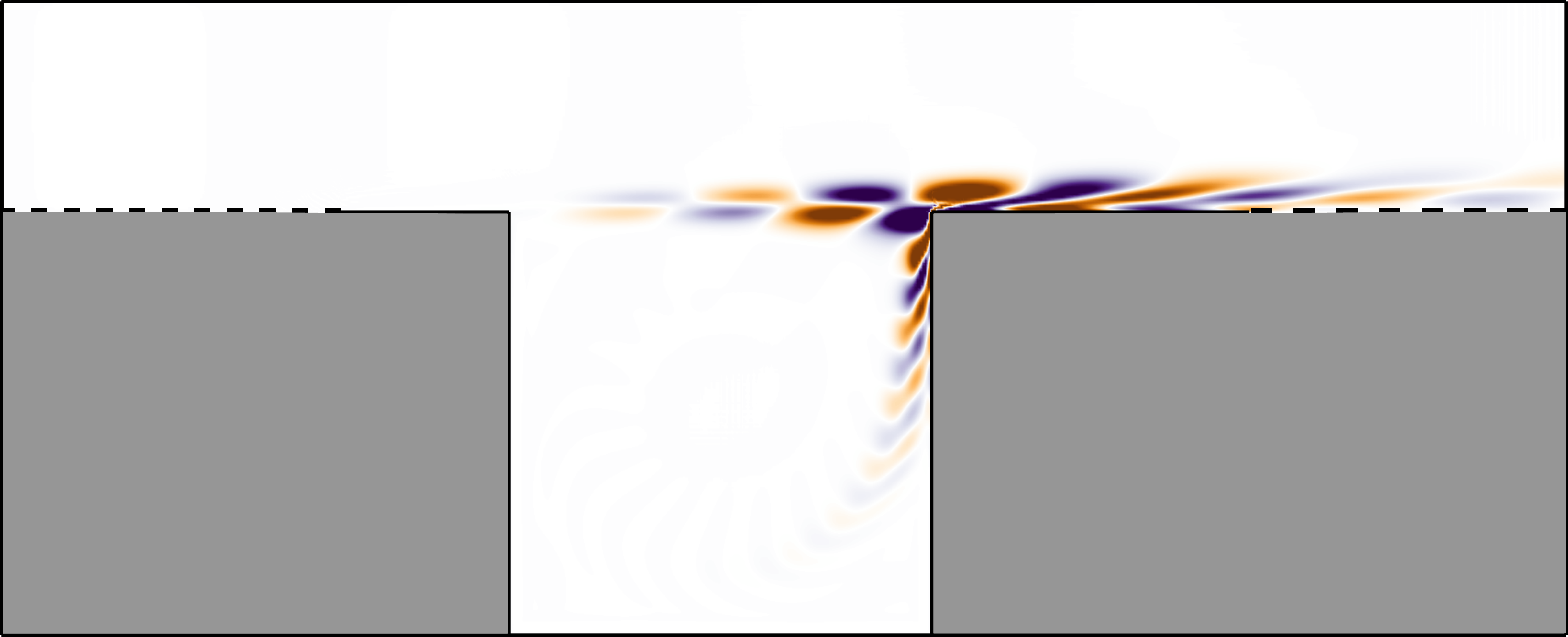}}
	\caption{(a) Eigenspectrum of the linearised Navier-Stokes operator for the shear-driven cavity flow at $\Rey=4250$. Vorticity fields of (b) the base flow and (c) the leading linearly unstable eigenmode. The dashed lines indicate the spatial extent over which the free-slip boundary condition is imposed.}
	\label{fig: cavity stability}
\end{figure}

\begin{figure}
	\centering
	\subfigure[]{\includegraphics[width=.45\textwidth]{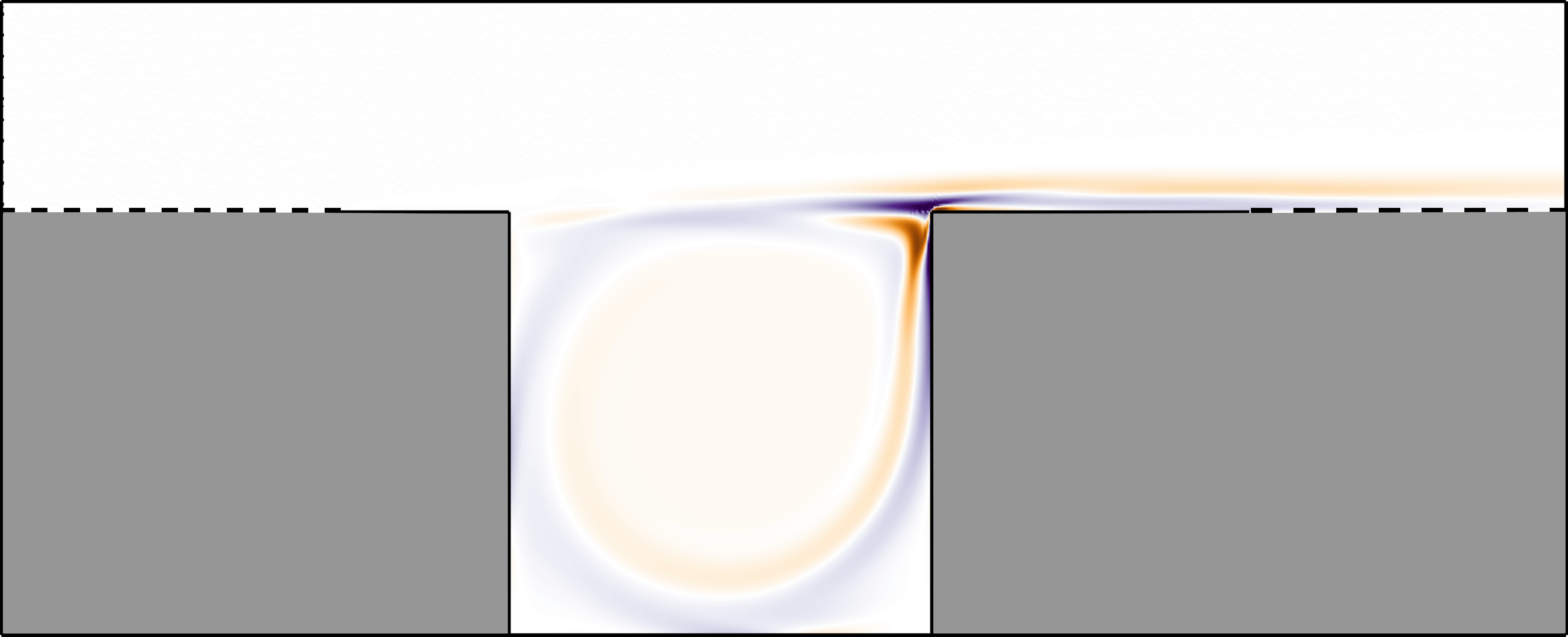}}
	\subfigure[]{\includegraphics[width=.45\textwidth]{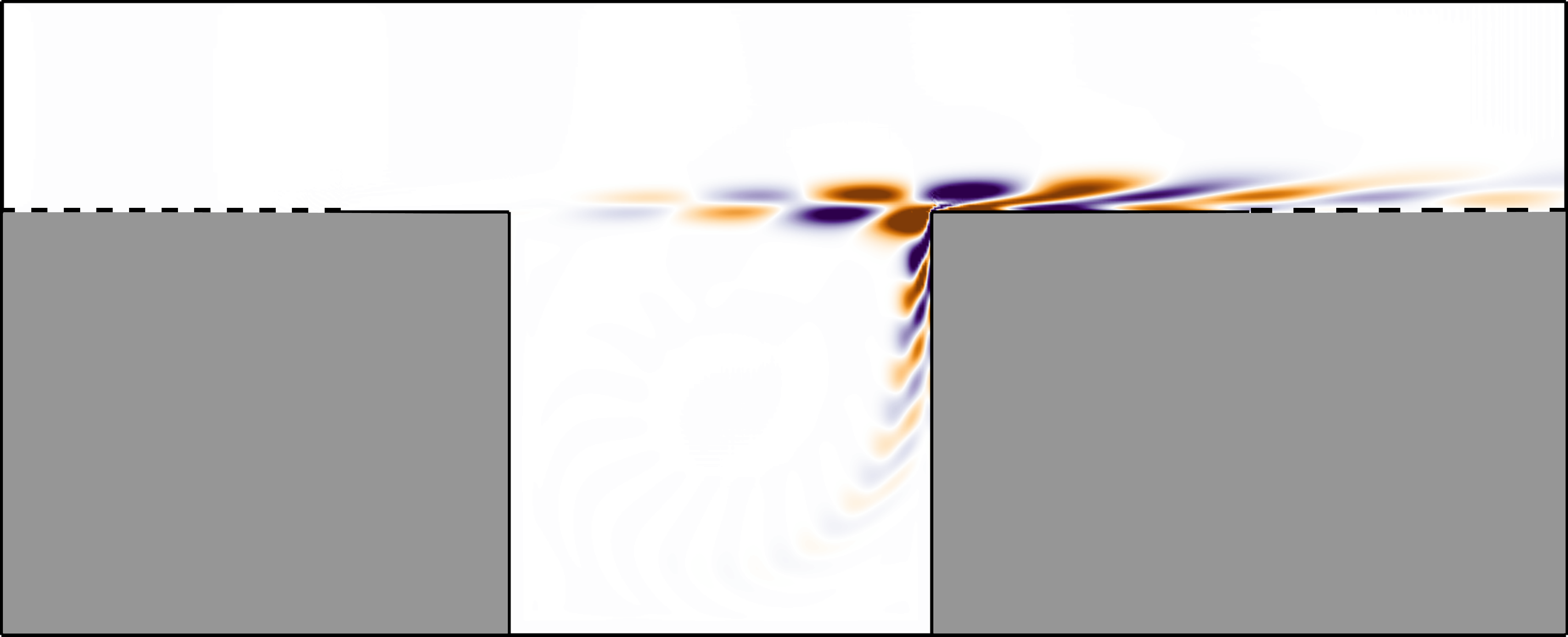}}
	\caption{Vorticity fields of (a) the shift mode and (b) the first POD mode for the shear-driven cavity flow at $Re=4250$.}
	\label{fig: cavity POD modes}
\end{figure}

\begin{figure}
	\centering
	\includegraphics[width=\textwidth]{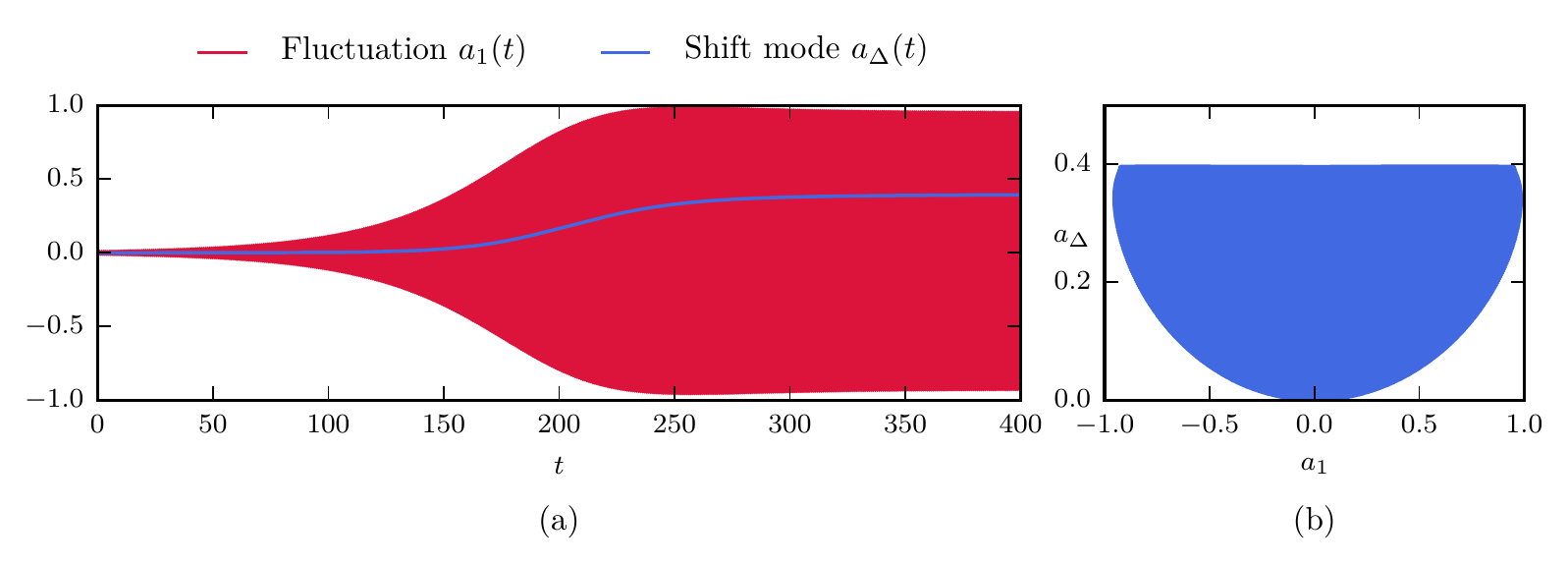}
	\caption{(a) Time evolution of the POD coefficients for the shear-driven cavity flow at $\Rey=4250$. The time evolution of $a_2(t)$, not shown, is very similar to that of $a_1(t)$. (b) Trajectory in the phase space projected onto the $a_1-a_{\Delta}$ plane. Note that, for both figures, the $a_i(t)$ coefficients have been multiplied by 100.}
	\label{fig: cavity dataset}
\end{figure}

    The computational domain and boundary conditions considered are the same as in \cite{jfm:sipp:2007}. The Reynolds number is set to $\Rey = 4250$, based on the free-stream velocity $U_{\infty}$ and the depth $L$ of the open cavity. As for the cylinder, the linearly unstable flow, the corresponding eigenspectrum and the vorticity field of the leading unstable eigenmode are presented in figure \ref{fig: cavity stability} for the sake of completeness. Using this linearly unstable flow as initial condition, another direct numerical simulation has been run until a statistically steady state is achieved. The vorticity field of the corresponding shift mode and of the first dominant POD mode are shown in figure \ref{fig: cavity POD modes}(a) and (b), respectively. While the leading unstable eigenmode and the dominant POD mode of the cylinder flow are extremely different, this is not the case for the shear-driven cavity flow at $\Rey=4250$. Comparing figure \ref{fig: cavity stability}(c) and figure \ref{fig: cavity POD modes}(b), it can be seen that these two modes are now very similar. The evolution in time of the coefficients $a_1(t)$ (dominant POD mode) and $a_{\Delta}(t)$ (shift mode) is shown in figure \ref{fig: cavity dataset}. Note that these curves appear as filled-in regions due to the high-frequency oscillations of $a_1(t)$. Despite the fundamental difference of the geometry, the different frequency of the oscillations and the smaller growth rate of the instability, the two flows considered herein appear to exhibit relatively similar dynamics when looking at the systems' trajectories projected onto the $a_1$-$a_{\Delta}$ planes: both low-dimensional representations of the flows appear to evolve along a parabolic manifold, see figure \ref{fig: cylinder dataset}(b) and figure \ref{fig: cavity dataset}(b).

\section{Results and discussion}
\label{sec: discussion}

	Following the seminal work of \cite{jfm:noack:2003}, so-called quadratic {\it Galerkin Regression models} are made of the basic building blocks necessary for reduced-order modelling of the flow configurations considered, {\it i.e.} a linear operator $\tilde{\mathcal{L}}$ and an energy-preserving quadratic nonlinearity $\tilde{\mathcal{N}}({\bf a})$. For that purpose, the library $\boldsymbol{\Uptheta} ({\bf a})$ used in the identification process is defined as $P_2({\bf a})$, {\it i.e.} all the polynomials of degree 2 or less in the entries of ${\bf a}$. The quadratic  Galerkin regression models identified for the cylinder flow and the shear-driven cavity flow are reported in tables \ref{tab: model A1} and \ref{tab: model A2}. A second type of models, cubic Galerkin Regression models, are made of the same basic building blocks as their quadratic counterparts. They moreover include higher-order nonlinearities which can serve to model the truncated modes, as discussed in \textsection \ref{subsec: higher-order constraints}. For that purpose, the library $\boldsymbol{\Uptheta} ({\bf a})$ used in the identification process is defined as $P_3({\bf a})$, {\it i.e.} all the polynomials of degree 3 or less in the entries of ${\bf a}$. Up to 57 coefficients then need to be identified for the present case with $n=3$ state variables. The cubic models identified for the cylinder flow and the shear-driven cavity flow are reported in tables \ref{tab: model B1} and \ref{tab: model B2}.

	\subsection{Cylinder flow}\label{Results:Cylinder}
	
\begin{figure}
	    \centering
	    \includegraphics[width=\textwidth]{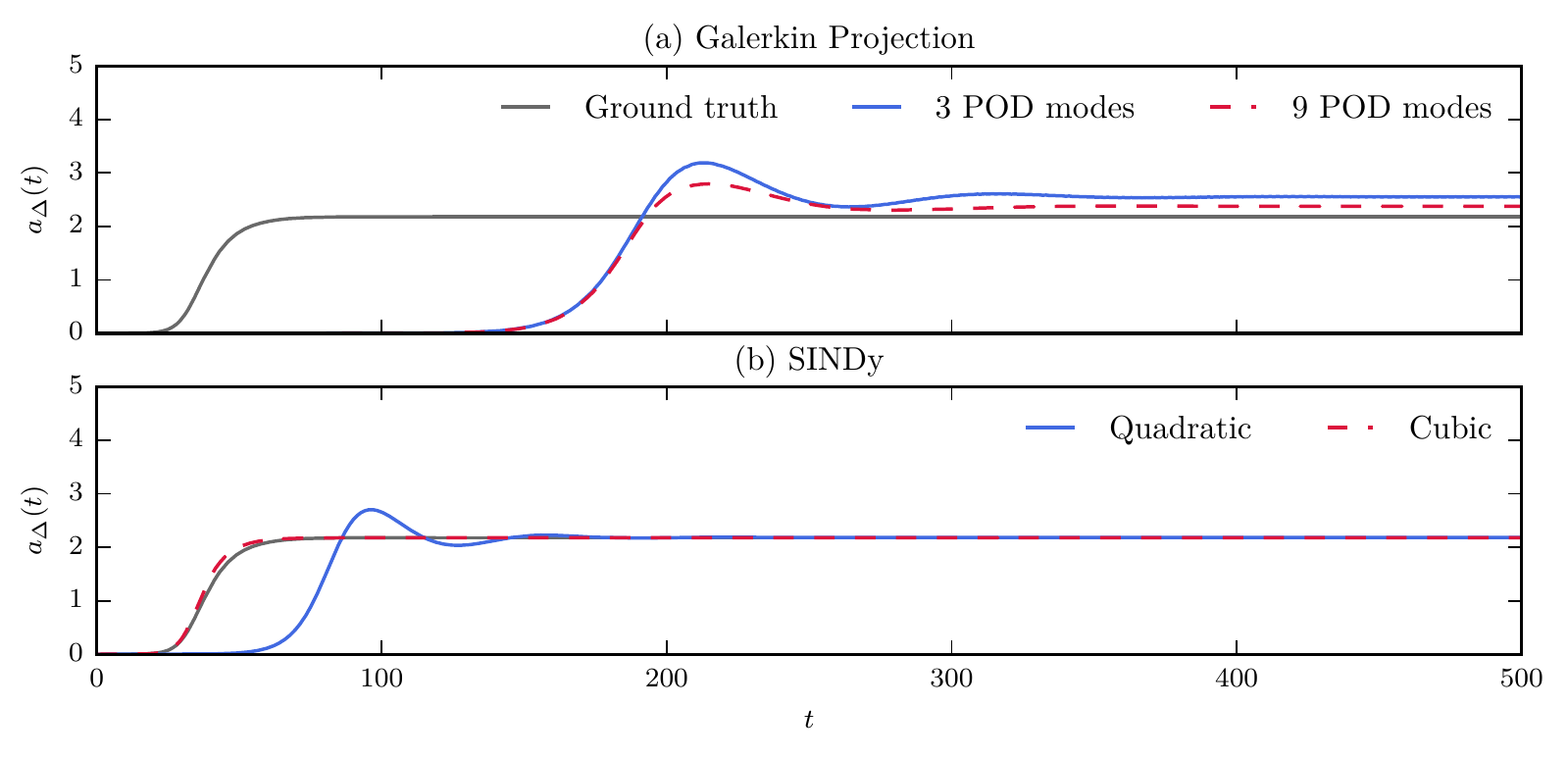}
	    \caption{Comparison of the time-evolution of the mean flow distortion $a_{\Delta}$ predicted by the different data-driven models for the two-dimensional cylinder flow at $\Rey=100$.}
	    \label{fig: comparison cylinder}
	\end{figure}

\begin{figure}
	    \centering
	    \includegraphics[width=\textwidth]{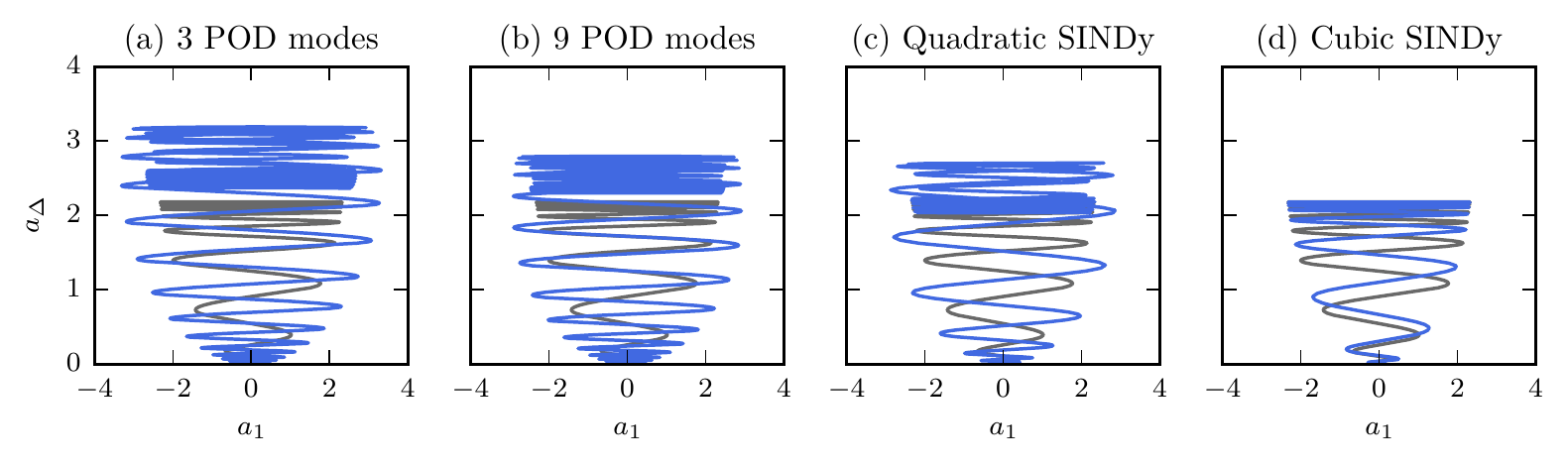}
	    \caption{Comparison of the trajectory in the $a_1-a_{\Delta}$ plane predicted by the different reduced-order models for the two-dimensional cylinder flow at $\Rey=100$. The light gray trajectory is the one given by a direct numerical simulation.}
	    \label{fig: comparison cylinder bis}
	\end{figure}

	Figures \ref{fig: comparison cylinder} and \ref{fig: comparison cylinder bis} provide a comparison of the dynamics predicted by the low-dimensional Galerkin Regression models identified using constrained sparse regression against the dynamics of the original system for the two-dimensional cylinder flow at $\Rey=100$. It also provides the dynamics predicted by two additional data-driven reduced-order models, namely:
	\begin{itemize}
	    \item the minimal Galerkin projection model including only the shift mode and the first two POD modes,
	    \item a Galerkin projection model including the shift mode and the first eight POD modes.
	\end{itemize}
	
    \bigskip
    
    Figure \ref{fig: comparison cylinder} depicts the evolution of the mean flow distortion as a function of time for the different reduced-order models. As reported in previous works \citep{jfm:noack:2003, arfm:rowley:2016}, the low-dimensional systems derived based on a Galerkin projection procedure that includes only the shift mode and the leading POD modes significantly over-estimate the duration of the transients. As explained by \cite{jfm:noack:2003}, this over-estimation results from the fact that the leading POD modes (see figure \ref{fig: cylinder POD modes}) provide only a crude approximation of the leading linear instability eigenmodes (see figure \ref{fig: cylinder stability}). These Galerkin projection models moreover suffer from an energy overshoot once nonlinear saturation kicks in. This overshoot and the ensuing larger amplitude of the mean flow distortion mostly result from the disruption of the energy cascade due to neglecting the higher-harmonic POD modes. Being neglected, these higher harmonics cannot absorb the excess energy produced by the two most energetic modes. The latter then grow beyond the correct value until the mean-flow distortion $a_{\Delta}(t)$ can eventually absorb this excess energy via the coupling terms. As shown in figure \ref{fig: comparison cylinder}(b), the quadratic {\it Galerkin Regression} model suffers from similar drawbacks, although the duration of transients is shortened and the final amplitude of the mean flow distortion is in agreement with that of the original system.
    
    \bigskip
    
    The dynamics predicted by the cubic Galerkin Regression model are shown in figures \ref{fig: comparison cylinder}(b) figure \ref{fig: comparison cylinder bis}(d). It can be seen that including higher-order nonlinearities results in a cubic Galerkin regression model that provides an almost perfect fit to the original data. The amplitude of the limit cycle is less than 0.5\% higher than that of the original system while the saturation of the mean flow distortion differs by less than 0.1\%. It has to be noted however that the growth rate of the instability is slightly over-estimated. Nonetheless, the inclusion of the cubic nonlinearities has a stabilising effect, hence preventing the energy overshoot and/or larger limit cycle amplitude observed for the quadratic models. A stabilising cubic term would also be obtained when deriving the Landau amplitude equation describing the transient dynamics of a small perturbation in the neighborhood of a supercritical Hopf bifurcation \citep{jfm:noack:1994, jfm:sipp:2007}. The excellent predictions of the cubic model as well as the existing connections with amplitude equations \citep{jfm:noack:1994, jfm:sipp:2007}, adiabatic elimination \citep{springer:haken:1983} or center manifold reduction \citep{book:wiggins:2003, jfm:carini:2015} thus justify {\it a posteriori} the use of higher-order nonlinearities to model the influence of the truncated modes onto the driving ones as discussed in \textsection \ref{subsec: higher-order constraints}.
    
	\subsection{Shear-driven cavity flow}
	
\begin{figure}
	    \centering
	    \includegraphics[width=\textwidth]{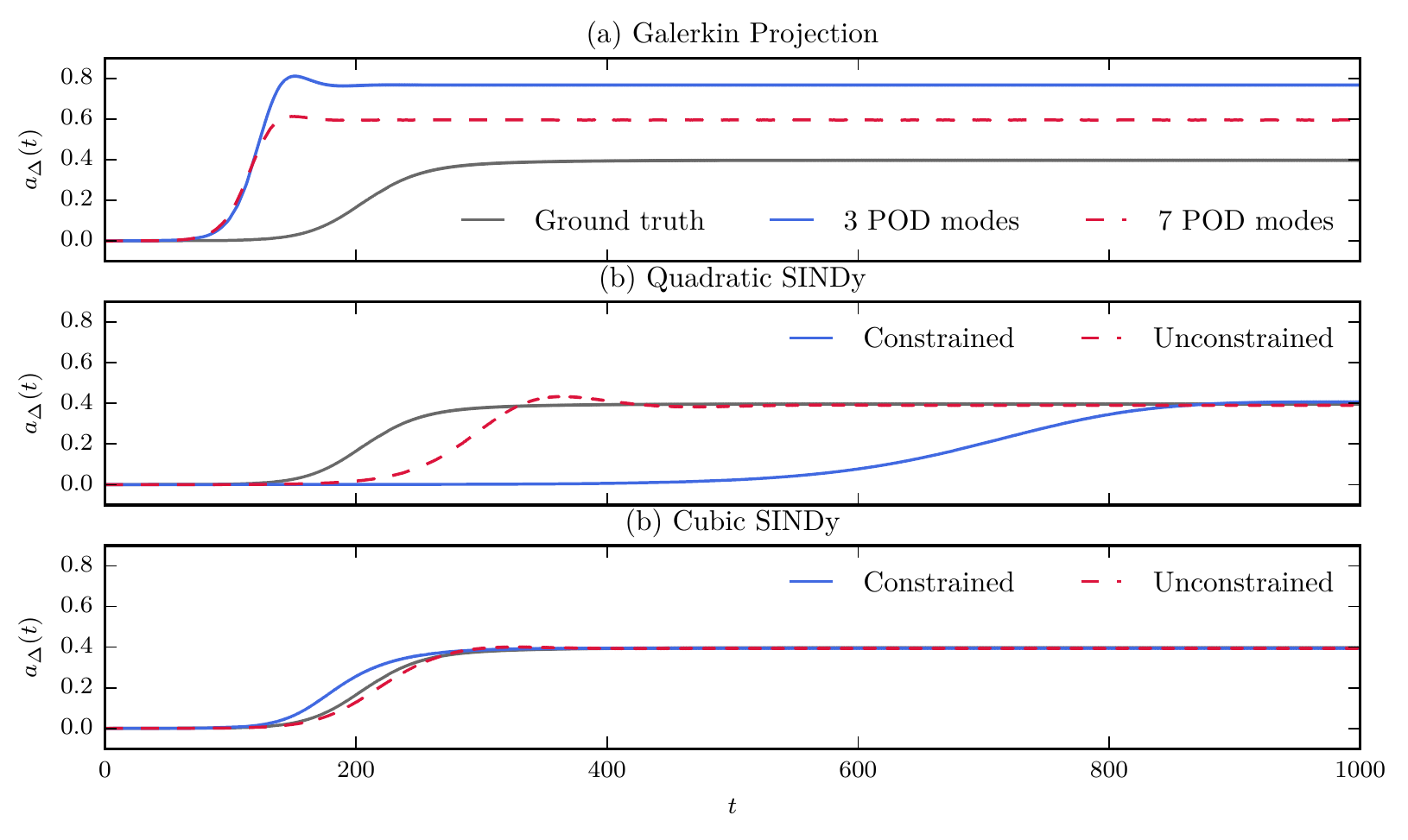}
	    \caption{Comparison of the time-evolution of the mean flow distortion $a_{\Delta}$ predicted by the different data-driven models for the two-dimensional shear-driven cavity flow at $\Rey=4250$.}
	    \label{fig: comparison cavity}
	\end{figure}

\begin{figure}
	    \centering
	    \includegraphics[width=\textwidth]{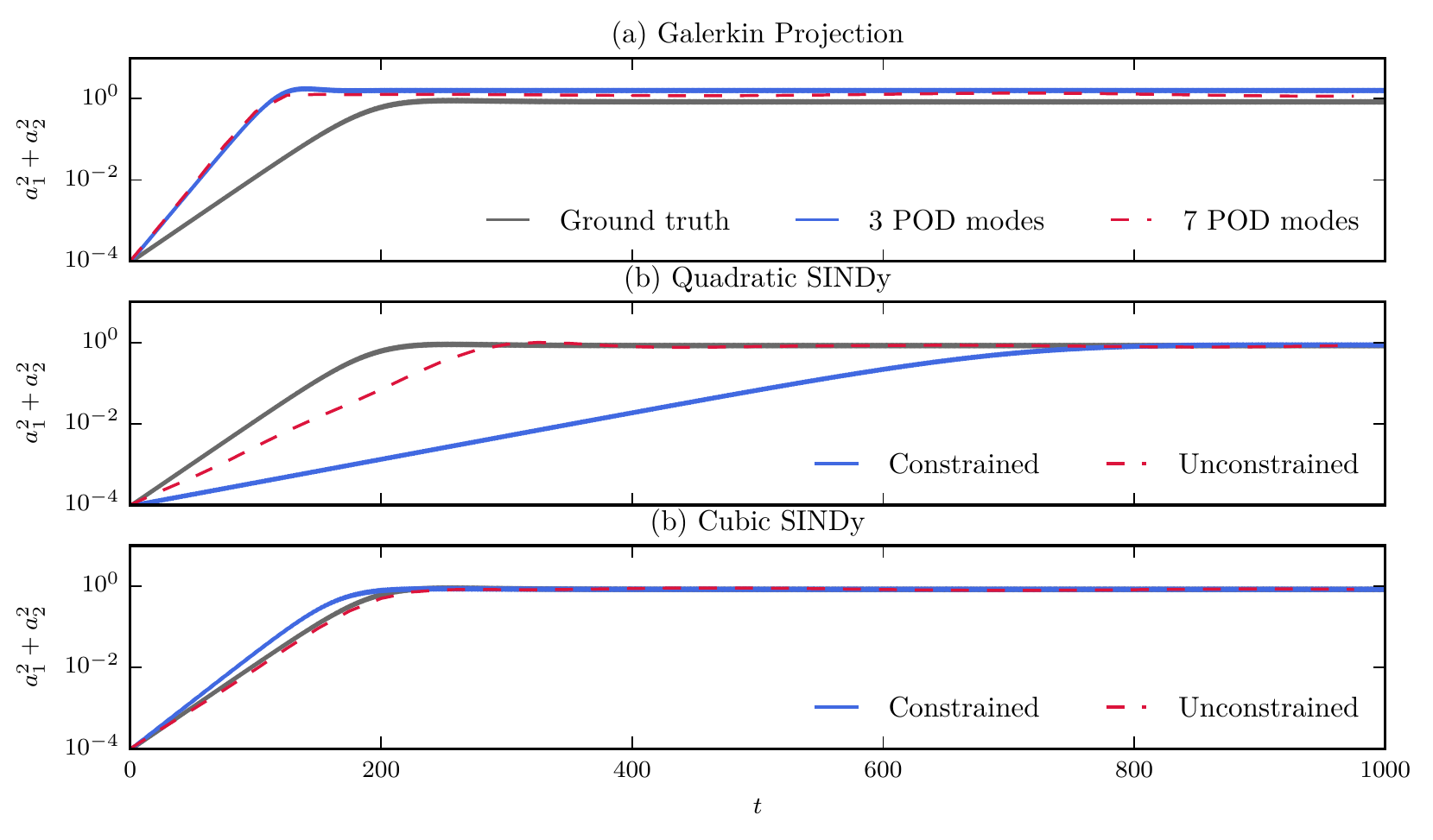}
	    \caption{Comparison of the time-evolution of the fluctuation's kinetic energy $a_1^2 + a_2^2$ predicted by the different data-driven models for the two-dimensional shear-driven cavity flow at $\Rey=4250$.}
	    \label{fig: comparison cavity bis}
	\end{figure}

	As for the cylinder flow, figures \ref{fig: comparison cavity} and \ref{fig: comparison cavity bis} provide a comparison of the dynamics predicted by the low-dimensional Galerkin Regression models identified using constrained sparse regression against the dynamics of the original system for the shear-driven cavity flow at $\Rey=4250$. The dynamics predicted by unconstrained SINDy models have also been included along with those predicted by the corresponding Galerkin projection models. Note that to identify meaningful Galerkin regression models, the sparse regression algorithm requires a pre-processing step so that all the features in $\boldsymbol{\Uptheta}({\bf a})$ have the same range in order to facilitate the optimisation procedure. 
	Although the geometry and the physics are quite different from that of the two-dimensional cylinder flow, it can be seen that the present \emph{Galerkin projection} models suffer from similar drawbacks as before: a misprediction of the transients duration and the saturation to higher mean flow distortion due to the disruption of the energy cascade. However, the key difference is that for the shear-driven cavity flow, the growth rate of the linear instability mode is slightly over-predicted by the Galerkin projection models.
	
	\bigskip
	
	Let us now draw our attention onto the quadratic Galerkin Regression models. Looking at the second subplot of figures \ref{fig: comparison cavity} and \ref{fig: comparison cavity bis}, it can be seen that both models correctly reproduce the asymptotic dynamics of the shear-driven cavity flow. The major difference however relies in the prediction of the transient dynamics. Although it would appear as the most physical one, the quadratic Galerkin regression model with an energy-preserving quadratic nonlinearity severely over-predicts the duration of the transients. Comparatively, the unconstrained quadratic model yields a much better prediction despite the small unphysical overshoot observed at the onset of nonlinear saturation. It is not clear at the present time why the constrained quadratic model performs so badly. One way to improve its performance would be to constrain the eigenspectrum of the low-dimensional linear operator to be a subset of its high-dimensional counterpart. Such a constraint on the determinant of the low-dimensional linear operator is however a non-convex constraint and does not fall in the scope of the library CVXOPT used in the present work.
	
	\bigskip
	
	Finally, it can be seen in figures \ref{fig: comparison cavity}(c) and \ref{fig: comparison cavity bis}(c) that both cubic models exhibit similar accuracy. The only visible difference between these two models is that the growth rate of the linear instability is slightly over-estimated by the constrained model while being slightly under-estimated by the unconstrained one. Given the similar performance, one might thus wonder what is the benefit of constraining the identification process. The answer to this question relies in the eigenspectrum of the low-dimensional linear operator. For the unconstrained model, this matrix and its eigenspectrum are given by
	
	\begin{equation}
	    \tilde{\mathcal{L}} = \begin{bmatrix}
	                            0 & 8.011 & 0.0408 \\
	                            -7.0579 & 0.0465 & -0.1146 \\
	                            -0.0181 & 0 & 0.0191
	                          \end{bmatrix} \text{ and } \Uplambda = \begin{bmatrix}                                                                 	                         0.0231 + i 7.519 & 0 & 0 \\
                                        	                           0 & 0.0231 - i 7.519 & 0 \\
                                                                       0 & 0 & 0.0194
                                         	                          \end{bmatrix}.
        \notag
	\end{equation}
	
	\noindent The unconstrained model hence correctly identifies the fixed point of the system as being a linearly unstable spiral within the $a_1-a_2$ plane. It also identifies it as being linearly unstable in the $a_{\Delta}$ direction. Naively, this result  appears consistent. Looking at the time-evolution depicted in figure \ref{fig: comparison cavity}(c) without prior knowledge of the problem, one could easily conclude that the system is linearly unstable in the $a_{\Delta}$ direction. From an identification point of view, the governing equations for $a_1$, $a_2$ and $a_{\Delta}$ are obtained independently from one another in the absence of constraints that would otherwise couple them. As a consequence, an equation predicting a linear instability of $a_{\Delta}$ is thus the simplest model identifiable which balances parsimony and consistency with observed measurements. However, given our prior knowledge about the physics of the problem, this is not an acceptable model. It could lead to a misunderstanding of the physics at play and seriously alter the practical performance of a linear or nonlinear controller based on such a faulty reduced-order model. As a comparison, the low-dimensional linear operator of the constrained cubic Galerkin regression model and the corresponding eigenspectrum are given by 
	
	\begin{equation}
	    \tilde{\mathcal{L}} = \begin{bmatrix}
	                            0.0063 & 8.2337 & 0.1442 \\
	                            -7.2843 & 0.049 & 0 \\
	                            -0.0347 & 0 & -0.0243
	                          \end{bmatrix} \text{ and } \Uplambda = \begin{bmatrix}                                                                 	                         0.0276 + i 7.75 & 0 & 0 \\
                                        	                           0 & 0.0276 - i 7.75 & 0 \\
                                                                       0 & 0 & -0.0243
                                         	                          \end{bmatrix}.
        \notag
	\end{equation}
	
	\noindent Given the linearly stable nature of the $a_{\Delta}$ direction now predicted, it is clear that coupling all of the equations governing the evolution of the system through the use of constraints mimicking the energy-preserving nature of the quadratic nonlinearity enables the identification of a much more physical low-dimensional system.

\section{Conclusion}
\label{sec: conclusion}

    This paper develops a new data-driven \emph{Galerkin regression} framework to identify nonlinear reduced-order models of a fluid. The resulting models incorporate a number of beneficial features of standard Galerkin projection, making them easy to interpret and use, but without the need for access to a high-fidelity Navier-Stokes model for the projection.  Galerkin regression models also provide a more flexible model identification, in that they readily generalize to include higher-order nonlinear terms that model the effect of truncated modes; the inclusion of these terms is shown to be extremely effective in the examples presented here. The Galerkin regression framework leverages the recent sparse identification of nonlinear dynamics (SINDy) algorithm \citep{pnas:brunton:2016}, and significantly generalizes it to include user-provided constraints directly into the sparsity-promoting regression. These additional constraints can be used to enforce \emph{a priori} known values of some of the regression coefficients, inherent symmetries of the system of equations or some physical behaviour such as the energy-preserving nature of the quadratic nonlinearity of the Navier-Stokes equations.
    
    \bigskip
    
    The Lorenz system, the two-dimensional cylinder flow and the shear-driven cavity have each been carefully analyzed to illustrate the system identification capabilities of the resulting algorithm. For that purpose, two polynomial libraries have been used and the constraints have been chosen in order to enforce different physical properties. The accuracy and performance of the so-called \emph{Galerkin regression} models have been compared against reduced-order models derived using a classical Galerkin projection method. All of the regression models qualitatively reproduce the main features of the original system: linear instability of the fixed point and final saturation to a periodic limit cycle. Though these models rely essentially on a data-driven approach, visual inspection of their trajectories in the phase space highlights the connection between the quadratic models and the models obtained using a Galerkin projection procedure in the seminal work of~\cite{jfm:noack:2003}. Moreover, both flow configurations highlight the importance of including cubic nonlinearities into the admissible pool of functions for the identification process, something utterly impossible with classical Galerkin projection without significant additional post-analysis. These cubic terms then model the influence of the truncated modes onto the driving ones, eventually enabling the identification of a low-dimensional system with much better predictive capabilities. Although the unconstrained cubic low-dimensional model of the shear-driven cavity reproduces faithfully the dynamics of the original system, this particular flow configuration has highlighted the importance of incorporating physically meaningful constraints into the regression to ensure that the identified model has the correct physical behaviour. In their absence, the SINDy algorithm incorrectly identifies the mean flow distortion as a linearly unstable manifold of the fixed point, while adding constraints results in the correct identification of a linearly stable eigenvalue.
    
    \bigskip
    
    Despite its promise, such an approach to system identification still suffers from certain limitations. One such limitation is illustrated by the quadratic constrained model identified for the shear-driven cavity flow which strongly under-estimates the growth rate of the linear instability. Given prior knowledge of the linear stability of the high-dimensional system, see \textsection \ref{sec: flow configurations}.2, one could then constrain the eigenspectrum of the low-dimensional linear operator to be a subset of its high-dimensional counterpart. Such a  constraint, involving the determinant of the low-dimensional matrix, falls outside the scope of convex optimisation. Current developments, based on the nonlinear optimisation library NLOPT~\citep{Johnson2014nlopt}, attempt to overcome such limitations. One might also argue that the systems considered in the present work are inherently low-dimensional and are thus not representative of the high-dimensionality of a transitional or turbulent flow. However, such flows have already been modelled with some success using a Galerkin projection procedure \citep{pof:gloerfelt:2008}. Given the parallels drawn in the present work between Galerkin projection and Galerkin regression, there is reason to believe that the present approach may be successfully applied to such flows as well. Indeed, this is an exciting future direction and is the subject of ongoing work. Including high-order nonlinear terms in the pool of admissible functions in combination with the sparsity-promoting capabilities of the algorithm might furthermore allow the identification of smaller and more robust reduced-order models without significantly altering their accuracy and predictive capabilities.
    
\section*{Acknowledgment}

We are grateful for many fruitful discussions with Bernd Noack, Josh Proctor and Nathan Kutz.  
We also appreciate valuable feedback from Scott Dawson and Clancy Rowley.  
SLB acknowledges generous funding support from the Defense Advanced Research Projects Agency (DARPA HR0011-16-C-0016) and from the Air Force Office of Scientific Research (AFOSR FA9550-13-1-0183).  

\appendix

\section{Coefficients of the different models identified}
\label{appendix: model coefficients}

The following tables provide the coefficients for each model identified using the SINDy algorithm extended with the energy-preserving constraint for the quadratic nonlinear term. Models A1 (see table \ref{tab: model A1}) and B1 (see table \ref{tab: model B1}) are the quadratic and cubic Galerkin regression models obtained for the two-dimensional cylinder flow at $\Rey=100$, respectively. Their counterparts for the shear-driven cavity flow, {\it i.e.} models A2 and B2, are given in tables \ref{tab: model A2} and \ref{tab: model B2}.

\begin{table}
	\centering
	\begin{tabular}{cccc}
		~ & $\dot{a}_1$ & $\dot{a}_2$ & $\dot{a}_{\Delta}$ \\
		$a_1$     &  0.0523 & 0.6667 & 0\\
		$a_2$     & -0.6856 & 0.0617 & 0\\
		$a_{\Delta}$     & 0           & 0          & -0.0513\\
		$a_1^2$ & 0           & 0          & 0.0245\\
		$a_1 a_2$   & 0           & 0          & 0\\
		$a_1 a_{\Delta}$   & -0.0245 & 0.1599 & 0  \\
		$a_2^2$ & 0          & 0           & 0.025  \\
		$a_2 a_{\Delta}$   & -0.1599 & -0.025  & 0       \\
		$a_{\Delta}^2$ & 0          & 0          & 0
	\end{tabular}
	\caption{Coefficients of the quadratic Galerkin regression model for the two-dimensional cylinder flow at $\Rey=100$.}
	\label{tab: model A1}
	
	\bigskip
	
	\begin{tabular}{cccc}
		~ & $\dot{a}_1$ & $\dot{a}_2$ & $\dot{a}_{\Delta}$ \\
		$a_1$                   &  $5.092 \cdot 10^{-3}$ & -7.068                        & 0\\
		$a_2$                   & 7.987                        & $8.166 \cdot 10^{-3}$ & 0\\
		$a_{\Delta}$         & $-2.369 \cdot 10^{-2}$ & 0.219                     & -0.034\\
		$a_1^2$               & 0                               & -0.1543                    & 0\\
		$a_1 a_2$             & 0.1542                     & 0.4106                     & 0\\
		$a_1 a_{\Delta}$   & 0                              & -0.0527                      & 0  \\
		$a_2^2$                & -0.4106                   & 0                              & 0.0343 \\
		$a_2 a_{\Delta}$   & 0.0527                       & -0.0343                    & 0       \\
		$a_{\Delta}^2$     & 0                               & 0    & 0
	\end{tabular}
	\caption{Coefficients of the quadratic Galerkin regression model A2 for the two-dimensional shear-driven cavity flow at $\Rey=4250$.}
	\label{tab: model A2}
\end{table}

\begin{table}
	\centering
	\begin{tabular}{cccc}
		~ & $\dot{a}_1$ & $\dot{a}_2$ & $\dot{a}_{\Delta}$ \\
		$a_1$                         &  0.0768  & 0.7527  & 0\\
		$a_2$                         & -0.745    & 0.1046  & 0\\
		$a_{\Delta}$               & 0            & 0           & -0.0357\\
		$a_1^2$                     & 0            & 0           & 0.0596\\
		$a_1 a_2$                   & 0           & 0           & 0\\
		$a_1 a_{\Delta}$         & -0.0596 & 0.1237  & 0  \\
		$a_2^2$                     & 0            & 0           & 0.0641  \\
		$a_2 a_{\Delta}$         & -0.1236 & -0.0641 & 0       \\
		$a_{\Delta}^2$            & 0           & 0           & 0 \\
		$a_1^3$                      & 0           & -0.0264 & 0 \\
		$a_1^2 a_2$                & 0.0318 & -0.005   & 0 \\
		$a_1^2 a_{\Delta}$      & 0          & 0           & -0.0189 \\
		$a_1 a_2^2$                & 0          & -0.0275 & 0 \\
		$a_1 a_2 a_{\Delta}$   & 0           & 0           & 0 \\
		$a_1 a_{\Delta}^2$      & 0.0107 & 0.025     & 0 \\
		$a_2^3$                      & 0.0323  & -0.005   & 0 \\
		$a_2^2 a_{\Delta}$      & 0           & 0          & -0.0208 \\
		$a_2 a_{\Delta}^2$      & -0.0358 & 0.0135 & 0 \\
		$a_{\Delta}^3$             & 0           & 0          & 0 \\
	\end{tabular}
	\caption{Coefficients of the cubic Galerkin regression model B1 for the two-dimensional cylinder flow at $\Rey=100$.}
	\label{tab: model B1}
	
	\bigskip
	
	\begin{tabular}{cccc}
		~ & $\dot{a}_1$ & $\dot{a}_2$ & $\dot{a}_{\Delta}$ \\
		$a_1$    								 &  $6.33 \cdot 10^{-3}$ 					& -7.284 								& -0.0347\\
		$a_2$     								& 8.233 												& 0.049 								& 0\\
		$a_{\Delta}$    					 & 0.144        									   & 0      								    & -0.0243\\
		$a_1^2$ 								& 0         											  & -0.0765   					      & 0\\
		$a_1 a_2$ 							  & 0.0765        								   & 0.088    					     & 0\\
		$a_1 a_{\Delta}$				   & 0 													& 13.53							 & 1.139  \\
		$a_2^2$ 								& -0.0881   									      & 0        						   & 0.0364  \\
		$a_2 a_{\Delta}$ 				  & -13.53 										& -0.036  							& -0.351       \\
		$a_{\Delta}^2$					 & -1.1393      						    & 0.351      							   & 0 \\
		$a_1^3$ 								& 0.0053											 & -2.219								 &-0.1659 \\
		$a_1^2 a_2$						 & 2.176											 & -0.063								 &0.063 \\
		$a_1^2 a_{\Delta}$		 & 0.0805											 & 0.549 								& 0 \\
		$a_1 a_2^2$					 & 0 														& -2.487								 &-0.184 \\
		$a_1 a_2 a_{\Delta}$ 	& 0.2242 											& -0.389								 &0 \\
		$a_1 a_{\Delta}^2$ 		& -0.0398 											& -20.499							 &-1.725 \\
		$a_2^3$							 & 2.445 												& -0.0592							 &0.0723 \\
		$a_2^2 a_{\Delta}$ 		& 0 														& 0.0602 							&-0.0195 \\
		$a_2 a_{\Delta}^2$ 		& 20.472											 & 0.0579								 &0.527 \\
		$a_{\Delta}^3$ 				& 1.805												 & -2.75								 & 0\\
	\end{tabular}
	\caption{Coefficients of the cubic Galerkin regression model B2 for the two-dimensional shear-driven cavity flow at $\Rey=4250$. }
	\label{tab: model B2}
\end{table}

    \section{Influence of the sparsity knob $\lambda$ and model selection}\label{AppendixB}

	Although it has not been discussed in the core of the present paper, the choice of the sparsity knob $\lambda$ is of crucial importance in the selection of the final model. Governing the level of sparsity, this parameter $\lambda$ is thus directly related to the accuracy and complexity of the identified models. If $\lambda$ is too small, very few terms are eliminated and the identified model has an artificially high complexity. On the other hand, if $\lambda$ is too large, the identified model may have too few terms, thus impacting its accuracy. To evaluate {\it a priori} the predictive capabilities of the identified model, it is convenient to analyze the number of non-zero coefficients and the $r^2$ score~\citep{book:draper:2014} (also known as the coefficient of determination) as a function of the sparsity knob $\lambda$. 
	
	\bigskip
	
\begin{figure}
    	\centering
    	\includegraphics[width=.9\textwidth]{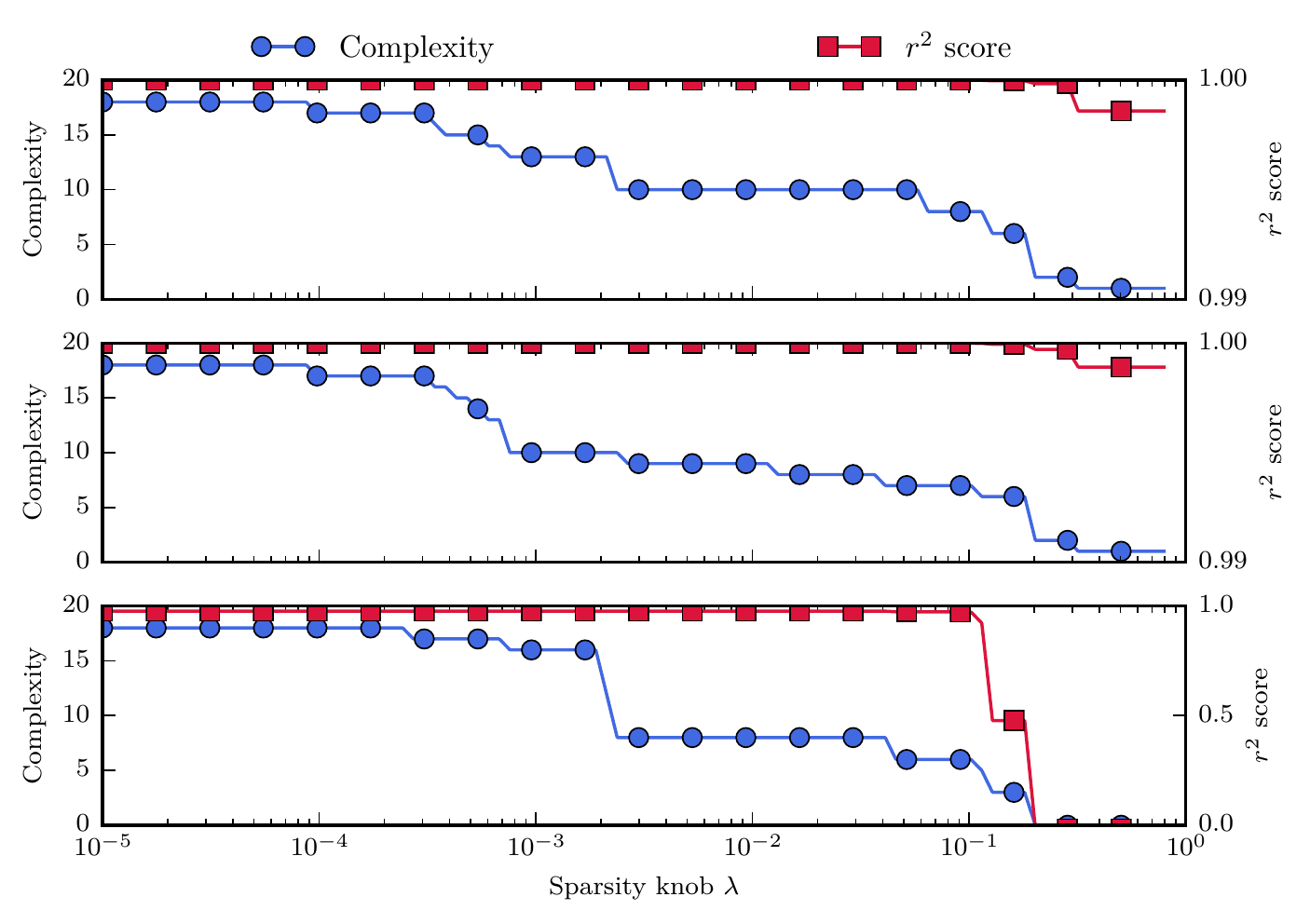}
    	\caption{Illustration of the influence of the sparsity knob $\lambda$ on the number of coefficients retained in the cubic SINDy model for the two-dimensional cylinder flow at $\Rey=100$. Evolution of the number of non-zero coefficients and of the $r^2$ score as a function of $\lambda$ for the governing equation of $a_1$ (top), $a_2$ (middle) and $a_{\Delta}$ (bottom).}
    	\label{fig: sparsity knob}
    \end{figure}

	Figure \ref{fig: sparsity knob} depicts the evolution of these two metrics as a function of the sparsity knob $\lambda$ for the cubic Galerkin regression model of the cylinder flow. It can be observe that increasing $\lambda$ up to almost 1 has a negligible influence of the {\it a priori} accuracy of the equations identified for the fluctuation's dynamics. Paradoxically, all the coefficients for the mean flow distortion's governing equation are set to zero for $\lambda > 0.1$, thus highlighting the over-aggressive sparsity promotion of the algorithm for this particular knob. This sudden drop in the $r^2$ score and number of non-zero coefficients of the mean flow distortion equation corresponds to the existence of a kink in the Pareto fronts of all three equations. The corresponding model is the one that provides the highest {\it a priori} accuracy while having the lowest complexity. All the identified models presented in this work have been selected using the same criterion. It has to be noted that such a model selection strategy can be combined with more sophisticated K-fold cross-validation to get an even better estimate of the {\it a priori} accuracy of the identified models.

\bibliography{bibliography}

\end{document}